\documentclass{emulateapj}
\usepackage{lscape}

\newcommand{\etal}{{et al.~}}
\def\teff{T$_{\rm{eff}}\,$}

\def\Dwa{$\,$\uppercase\expandafter{\romannumeral5}$\,$}
\def\mic{$\mu$m$\,$}

\def\sless{\lower2pt\hbox{$\buildrel {\scriptstyle <}
   \over {\scriptstyle\sim}$}}
\def\sgreat{\lower2pt\hbox{$\buildrel {\scriptstyle >}
   \over {\scriptstyle\sim}$}}

\begin{document}

%\pagenumbering{arabic}

\title{{\it Spitzer}/IRAC Photometry of M, L, and T Dwarfs}
\shorttitle{IRAC Photometry of M, L, and T Dwarfs}
\shortauthors{Patten et al.}

\author{Brian M.\ Patten\altaffilmark{1}$^,$\altaffilmark{2}, 
John R.\ Stauffer\altaffilmark{3}, 
Adam Burrows\altaffilmark{4},  
Massimo Marengo\altaffilmark{1}, 
Joseph L.\ Hora\altaffilmark{1},
Kevin L.\ Luhman\altaffilmark{1}$^,$\altaffilmark{5}, 
Sarah M.\  Sonnett\altaffilmark{1}$^,$\altaffilmark{6}, 
Todd J.\ Henry\altaffilmark{7},  
Deepak Raghavan\altaffilmark{7},
S.\ Thomas Megeath\altaffilmark{1}, 
James Liebert\altaffilmark{4},   
\& Giovanni G.\ Fazio\altaffilmark{1}}

\altaffiltext{1}{Harvard-Smithsonian Center for Astrophysics, 60 Garden St. Cambridge, MA 02138-1516} 
\altaffiltext{2}{bpatten@cfa.harvard.edu}
\altaffiltext{3}{{\it Spitzer} Science Center, Mail Stop 314-6, California Institute of Technology, Pasadena, CA 91125}
\altaffiltext{4}{Department of Astronomy, The University of Arizona, Tucson, Arizona, USA}
\altaffiltext{5}{Department of Astronomy and Astrophysics, Pennsylvania State University, 525 Davey Lab, University Park, PA USA 16802} 
\altaffiltext{6}{Department of Physics and Astronomy, College of Charleston, 101 Hollings Science Center, 58 Coming St., Charleston, SC USA 29424} 
\altaffiltext{7}{Department of Physics and Astronomy, Georgia State University, Atlanta, GA 30302-4106}

\begin{abstract}
We present the results of a program to acquire photometry for 
eighty-six late-M, L, and T dwarfs using the Infrared Array Camera (IRAC) 
on the {\it Spitzer Space Telescope}.  
We examine the behavior of these cool dwarfs in various color-color
and color-magnitude diagrams composed of near-IR and IRAC data.
The T dwarfs exhibit the most distinctive positions in these diagrams.
In M$_{5.8}$ versus [5.8]-[8.0], the IRAC data for T dwarfs are not monotonic
in either magnitude or color, giving the clearest indication yet that
the T dwarfs are not a one parameter family in \teff. Because metallicity 
does not vary enough in the solar neighborhood to act as the second 
parameter, the most likely 
candidate then is {\it gravity}, which in turn translates to 
{\it mass}.  Among objects with similar spectral type, the range of mass 
suggested by our sample is about a factor of five ($\sim$70 $M_{Jupiter}$ 
to $\sim$15 $M_{Jupiter}$), with the less 
massive objects making up the younger members of the sample.  We also 
find the IRAC 4.5 \mic fluxes to be lower than expected, from which
we infer a stronger CO fundamental band at $\sim$4.67 \mic.
This suggests that equilibrium CH$_4$/CO chemistry underestimates the 
abundance of CO in T dwarf atmospheres, confirming earlier results
based on $M$-band observations from the ground.  In combining 
IRAC photometry with near-IR $JHK$ photometry and parallax data, we
find the combination of $K_S$, IRAC 3.6 \mic, and 4.5 \mic bands
to provide the best color-color discrimination for a wide range 
of M, L, and T dwarfs.  Also noteworthy is the M$_{K_S}$ versus $K_S$-[4.5]
relation, which shows a smooth progression over spectral type and 
splits the M, L, and T types cleanly. 
\end{abstract}

\keywords{infrared: stars -- stars: fundamental properties -- stars: late-type -- stars: low-mass, brown dwarfs}

\section{Introduction}

The study of low-mass dwarfs has progressed enormously in the last decade: 
from the strictly theoretical beginnings for objects of sub-stellar mass 
(e.g. Kumar 1963; Grossman 1970; Nelson, Rappaport, \& Joss 1985) to the 
actual discovery in the late-1980's and mid-1990's of objects in two new 
spectral classes later than M (Becklin \& Zuckerman 1988; 
Nakajima et al. 1995).
As a result of a large amount of observational effort over the past decade, 
the temperature sequence of dwarfs has been extended down to T $\sim$ 700 K
(Golimowski et al. 2004a).  
Several hundred L dwarfs and several dozen T dwarfs have now been identified, 
leading to well-defined L and T spectral sequences (Kirkpatrick 2005), and 
to a large body of accurate photometry and quantitative spectroscopy to 
characterize those spectral types (Leggett et al. 2001, 2002; Basri et al. 
2000; Reid et al. 2002).  Also of importance, other groups have been able 
to obtain accurate trigonometric parallaxes for a large number of these 
very cool dwarfs (e.g., Dahn et al. 2002; Tinney et al. 2003; Vrba \etal 2004) 
and to obtain 
dynamical masses for a significant sample of stars (e.g., Henry \& McCarthy 1993; 
Henry et al. 1999; 
Lane \etal 2001; Zapatero Osorio \etal 2004; Bouy \etal 2004; 
Close et al. 2005).

The observational progress was accompanied by an equally dramatic 
evolution in the sophistication of the theoretical models of cool dwarfs.
The improved observational data spurred theoretical model efforts 
such as Allard et al. (1994) and Brett (1995), who were able to incorporate 
much more extensive molecular line lists 
and thus improve the fit to observations 
and extend the model spectra further into the infrared.  The model
atmospheres have since rapidly evolved, incorporating ever-larger molecular 
line lists, dust formation and dust-settling, and improved treatment of 
pressure broadening, particularly for the alkali lines that dominate the 
optical and near-IR spectra of the L dwarfs. More recently, the models
have begun to take into account non-equilibrium chemistry (e.g., 
Noll, Geballe, \& Marley 1997; Saumon et al. 2000; 
Burrows, Marley, \& Sharp 2000; 
Ackerman \& Marley 2001; Allard et al. 2001; Lodders \& Fegley 2002; 
Marley et al. 2002; Tsuji 2002; 
Saumon et al. 2003, 2006).   

While much of the work on sub-stellar mass objects to date has been carried 
out in the optical and near-IR, there is great potential to gain even more 
insight into brown dwarf atmospheres by extending observations into the 
mid-IR regime.  The advantages include the peaking of the spectral 
energy distribution for L and T dwarfs between 1$-$4 \mic and the 
presence of prominent molecular features such as CH$_4$, H$_2$O, and 
NH$_3$ 
(e.g. Burrows et al. 1997; 2001).  Ground-based infrared observations are hampered 
longward of $K$ band by telluric atmospheric lines, such as OH and H$_2$O, 
making the atmosphere increasingly opaque at these wavelengths.  
Thermal emission of the atmosphere and telescope structure also cause 
very high backgrounds for ground-based observations longward of $K$ band.
Thus, space-based instruments with sensitivity in the mid-infrared are 
required for the study of sub-stellar mass objects in this wavelength regime.  
It was in this context that the Infrared Array Camera (IRAC) Guaranteed 
Time Observer
(GTO) team decided that it would be important to obtain accurate mid-IR 
fluxes for a representative sample of M, L, and T dwarfs when the 
{\it Spitzer Space Telescope} was launched.

We present in this paper the results of this survey.  In section 2, 
we describe the sample selection of M, L, and T dwarfs and their basic 
properties.  In section 3, we summarize the observing strategy used 
with {\it Spitzer}/IRAC and the subsequent data reduction and extraction
of the IRAC photometry.  In section 4, we present the basic color-color 
and color-magnitude diagrams for both IRAC and near-IR $JHK$ photometry 
and discuss trending with color and spectral type.  Finally, in section
5, we conclude with a few tentative interpretations of the observations.

\section{The Sample}

The targets in our program were selected from the literature.
Because the GTO program target selection was frozen a year prior 
to {\it Spitzer} launch on 25 August 2003, our selection was limited 
to those objects cataloged at that time.  The primary
consideration that governed our selection process was whether 
or not the target had a measured trigonometric parallax.  The secondary
constraints on our selections were:  (1) we wanted
the sources to be relatively bright in order to ensure good 
signal-to-noise ratio on the photometry given our observing strategy
(see section 3), since these objects would represent the fiducial 
sample for other science programs, (2) the targets needed to have 
well-determined spectral types because of the desire to have 
representatives of all spectral sub-type bins from late-M through 
T types, (3) the targets needed to be located in relatively 
uncrowded fields, and (4) should not be close binaries (separations of
$\leq$ 6 arcseconds or about 5 IRAC pixels).  While we tried to 
avoid including
binaries in our sample, the information available in the literature
was/is quite incomplete, so some now-known binaries were included in 
our program.  Other objects may be binaries, but are currently 
not known as such.  Given that dwarfs in the L7$-$T2 spectral 
type range have been found to have a high binary fraction
(e.g. Burgasser et al. 2006), it seems inevitable that more objects in our 
sample will turn out to be binaries than we have currently noted.

The basic properties of the objects in our sample are summarized in
Table 1.  We provide here position, nomenclature, parallax (if available), 
spectral type, and near-IR photometry.  Some target names are abbreviated 
from their official forms.  Objects from the DENIS, 2MASS, and SDSS have had
their survey acronyms shortened to three characters - DEN, 2MA, and SDS
respectively - followed by the first four digits of right ascension and
then the first four digits of the declination.  
For the $JHK_S$ photometry, we use data from 2MASS primarily in order to 
maintain consistency.  However in those cases where the 2MASS errors are
large ($\ga$ 0.20 mag) or other flags in the database (e.g. blending,
contamination, confusion) suggest questionable photometric reliability,
we tabulate photometry from the literature (see references in Table 1).
Because the near-IR photometry for low-mass dwarfs is strongly 
system dependent (particularly for T dwarfs), we have chosen to rely
on MKO-system photometry when reliable 2MASS photometry is not 
available.  We make this choice because there are well documented 
transformations between these two systems (Stephens \& Leggett 2004)
as well as a wealth of MKO-system photometry for low-mass brown dwarfs
available in the literature (see additional discussion in section 4.1.2).
For two T dwarfs in our sample (SDS1624+0029 and 2MA1237+6526) we acquired 
new $JHK_S$ photometry with the PAIRITEL facility on Mt. Hopkins (telescope 
and instrument formerly used for the 2MASS project, see Bloom et al. 2006
for additional details).  These photometry were calibrated to the 2MASS
system using local comparison stars in each field-of-view. 
Because the T dwarfs are intrinsically faint in the optical,
it is difficult to classify these objects using the same criteria/features 
that 
work for the brighter L dwarfs.
For the spectral types in Table 1, we have chosen to quote optical types
for the M and L dwarfs while, for the T dwarfs, infrared types are used.  
Near-IR classification systems take advantage 
of the peak of the spectral energy distribution being located in the
infrared for T dwarfs and a wealth of molecular features in this 
wavelength regime.  The unified near-IR classification
system for T dwarfs (Burgasser et al. 2006) is compatible with 
optical classification systems (e.g. Kirkpatrick et al. 1999, 2000)
at type L8, creating a smooth continuum of types from M through T.
It is worth noting, however, that for the L dwarfs, the optical and near-IR 
spectral types can be quite different, as they probe different optical
depths in each wavelength regime (see discussion in Knapp et al. 2004;
Kirkpatrick 2005).

In addition to our primary MLT sample, unsaturated photometry was 
secured for a number of late-M dwarfs from a related GTO program
(PID 33 ``A Search for Companions Around Stars Within Five Parsecs'').
These include GJ 1002, LHS 288, GJ 412B, GJ 1111, LHS 292, 
SO0253+1652, and LP944-020.  Archival data from an unpublished {\it Spitzer} 
Early Release Observation yielded photometry for the T dwarf binary
$\epsilon$ Indi BC, which are also presented in Table 1.  Although
GJ 229B was included in the nominal MLT GTO program, it was lost in
the glare from GJ 229, so IRAC photometry could not be secured 
for this object.  In total, supporting data for eighty-six objects 
for which we were able to secure IRAC photometry are presented in Table 1.

\subsection{Comments on specific sources}

As of the writing of this paper, the following objects have been determined 
to be binaries, unresolved or blended at the IRAC plate scale of $\sim$1.2
arcseconds/pixel:  
GJ 1001BC, SDS0423-0414, 2MA0746+2000, GJ 337CD, SDS0926+5847, 
SDS1021-0304, Kelu-1, 
2MA1225-2739, 2MA1534-2952, 2MA1553+1532,
and 2MA1728+3948 (Reid \etal 2001; 
Burgasser \etal 2003a;
Gizis \etal 2003;  
Bouy et al. 2004; 
Golimowski \etal 2004b; 
Vrba \etal 2004; 
Burgasser, Kirkpatrick, \& Lowrance 2005; 
Burgasser \etal 2005;
Liu \& Leggett 2005;
Burgasser \etal 2006).
The following objects have been noted to have spectral peculiarities
-- unusually weak metal lines for their subtype, emission lines in their
spectra, or unusual colors -- suggesting that some of 
these may be metal-poor subdwarfs, however, 
multiplicity may also play a role: 2MA0559-1404, 2MA0937+2931, 
2MA1047+2124, SDS1110+0116, 2MA1217-0311, 2MA1237+6526, and SDS1624+0029 
(Burgasser \etal 2002;
Burgasser \etal 2003b;
Tinney \etal 2003;
Golimowski \etal 2004a).  

{\it 2MA0532+8246}: This object, the first identified late type L subdwarf
(Burgasser et al. 2003a), has been confirmed as such in 
Reiners \& Basri (2006).

{\it BRI0021-0214}: Photometric variability has been reported in $I$-band 
with a full amplitude about 0.05 mag and a possible periods of 
about 0.2 and 0.83 days (Mart\'in \etal 2001).  However, its large $v$ sin $i$
of $\sim$34--42 km/s (Tinney \& Reid 1998; Mohanty \& Basri 2003), 
if true, would indicate a rotation period significantly less than 0.8 days.   

{\it DENIS0255-4700}: Has a very high $v$ sin $i$ of 40 km/s (Mohanty \& Basri 
2003).  A suggestion of photometric variability could be explained 
in part by rotational modulation (Koen 2005).

{\it LP944-020}: Listed as member of the Castor moving group by
Ribas (2003), thus it's age is $\sim$300--500 Myr.
Tinney (1998) gives an age of 475--650 Myr for this star, 
based on the detection of lithium and the star's $L_{bol}$.  

{\it 2MA0451-3402}: Observed to have periodic photometric variability
with a period of $\sim$6.9 hours (Koen 2004).

{\it 2MA0746+2000AB}: Reported to be photometrically variable with a period
of 31 hrs according to Gelino (2002).  Its measured
$v$ sin $i$ (Bailer-Jones 2004) is inconsistent with
that rotation period, however.

{\it 2MA2224-0158}: Noted as having an anomalously red spectrum,
possibly indicative of an unusually thick condensate cloud
or of low surface gravity (Cushing et al. 2005).

\section{Observations}

IRAC is a four-channel camera that obtains simultaneous broadband images 
at 3.6, 4.5, 5.8, and 8.0 \mic (sometimes referred to as {\it channels} 
1$-$4 respectively). 
Two nearly adjacent 
5.2\arcmin~ $\times$ 5.2\arcmin~ fields of view in the focal plane (for
256 $\times$ 256 pixel detector arrays) are viewed by the four channels 
in pairs (3.6 and 5.8 \mic; 4.5 and 8 \mic).  Additional details 
on the design and performance of IRAC can be found in Fazio et al. (2004).

Profiles of the IRAC filters are shown in Figure 1.  For comparison,
the profiles of the MKO $L'$ and $M'$ filters, as well as a model spectrum
of a mid-T dwarf, are also shown to illustrate the 
relative band centers and bandwidths of filters in recent use to 
characterize the infrared colors of M, L, and T dwarfs 
(e.g., Leggett et al. 2002; Golimowski et al. 2004a; Knapp et al. 2004).  While 
the IRAC 3.6 \mic channel does share some similarities with the 
standard $L$ filter, it is important to note that the 3.6 and 4.5 \mic 
IRAC channels are significantly bluer than the $L'$, $M$, and $M'$ filters 
and/or are much broader than all of these filters.  The photometry presented 
in this paper have been calibrated using a zero magnitude defined by
Vega in all IRAC channels (e.g. Cohen 2003; Cohen et al. 2003; 
Reach et al. 2005).  As illustrated in Figure 1, the spectra of low and 
sub-stellar mass objects have a great deal of structure within the filter 
bandpasses.  Therefore, particularly for T dwarfs, IR fluxes measured
using broadband filters are critically dependent on both the 
central wavelength and the filter transmission profile.
Significant color-terms can be present when transforming fluxes 
measured in apparently similar filters from two different 
systems (e.g., from $L'$ to IRAC 3.6 \mic).

\subsection{Observing Strategy}

All observations in this GTO program share a common {\it Spitzer} Astronomical 
Observation Request (AOR) design.  Each target was observed using the 
same camera modes and frame times as well as dithering the targets over 
the same pixels (approximately) on the arrays. The primary motivation
of this strategy was to minimize the introduction of additional 
uncertainties into the relative photometry for each object.
Specifically, the AOR uses a 5-position Gaussian dither pattern,
starting with the target near the center of the array.  Dithering
mitigates the effects of cosmic rays and bad/hot pixels, while the small 
scale factor option for the pattern keeps the target near the center 
of the detectors in order to minimize spatially dependent uncertainties 
in the calibration of the instrument.  The relative offsets for each 
dither position in the pattern were identical for all of the objects
in the program and were distributed within a radius of 38 arcseconds
of the center/initial position.  High dynamic range mode was used because 
the program objects cover a wide range of brightness and the relative 
sensitivities of the four IRAC detectors preclude a single exposure time 
that will produce good signal-to-noise and unsaturated photometry in all 
four channels (see also discussion in section 3.3).  The frame times of 
30 and 2 seconds yield effective exposure 
times for each dither position of 26.8 and 1.2 seconds respectively.

The individual observations themselves were scheduled by the {\it Spitzer} 
Science Center (SSC) and were executed over the period 06 December 2003 to 
01 November 2004.  The data for each target, as received from the spacecraft, 
were processed by the IRAC pipeline software at the SSC. 
This pipeline
removes the electronic bias, subtracts a dark sky image generated from 
observations of relatively empty sky near the ecliptic pole, flat-fields 
the data using calibration observations of relatively blank fields near
the ecliptic plane, and then linearizes the data using laboratory 
measurements of each pixel's response to a calibration lamp in frames of
varying length\footnote{Additional details can be found in the 
IRAC Data Handbook at http://ssc.spitzer.caltech.edu/irac/data.html}.  
The absolute calibration of IRAC is derived from aperture photometry of 
standard stars.  This calibration is applied to the data such that the 
final pipeline product for each frame is in units of surface flux per 
steradian (i.e., Reach et al. 2005). The pipeline processing produces 
calibrated 
data for each frame in the dither pattern for each IRAC channel.  This 
constitutes the Basic Calibrated Data (BCD) that were used for our analysis.

\subsection{Photometry}

Photometry was extracted for each source using the aperture photometry 
package in IRAF.  To ensure the highest signal-to-noise and to minimize 
contamination by cosmic rays, the individual images in the dither pattern
in each channel were combined using a local post-pipeline software suite
developed by one of us (BMP).  This software was used to co-register the
frames in sky coordinates and then co-add the frames while rejecting 
temporally transient events (e.g., cosmic rays) as well as fixed-pattern
noise and bad pixels.  We note that caution must be used with any kind 
of filtering scheme to remove transients when co-adding or mosaicking 
IRAC data.  Because the point 
spread function (PSF) is under/critically-sampled by the IRAC arrays, 
this can lead to a pixel response function (PRF) that is sharply
peaked if the PSF is centered on a pixel or to a PRF with a broader,
lower-intensity peak if the PSF is centered at the corners of four 
adjacent pixels (the two extreme cases).  Thus, the pixel phase 
(i.e., the position within a pixel) of the PSF centroid becomes important 
when using transient rejection based on, say, median-filtering and/or
sigma clipping.  The large changes in the peak of the PRF for a 
dithered source could lead to real flux being rejected and, thus,
to an artificial reduction in the count rate of the source when 
combining the individual frames of the dither sequence (see 
additional discussion in Schuster, Marengo, \& Patten 2006).

To measure the photometry, we first removed the surface flux calibration 
from the images by dividing by the flux conversion factor (MJy/sr per 
DN/sec) found in the image headers of the BCD data and by multiplying by 
the effective integration time - 26.8 or 1.2 seconds for the individual 
frames in the dither pattern and 134 or 6 seconds for the co-added, cleaned 
data.  The IRAC absolute calibration is based on observations of standard 
stars measured with aperture photometry using a source aperture with 
a radius of 10 native IRAC pixels in each channel.  The background was 
estimated using an annulus centered on the source position with an
inner radius 10 pixels and width of 10 pixels (Reach et al. 2005).  
Because many of the targets in our program are located in semi-crowded 
fields, we chose to use a smaller source aperture with a radius of 4 native 
IRAC pixels in order to avoid contaminating flux from other, nearby sources.  
An additional benefit of using a smaller source aperture is an improved 
signal-to-noise ratio for many of the fainter sources.  For the background
estimation, we used the same annulus as that defined for the IRAC absolute
calibration.  Photometry was extracted from both the co-added frames and the 
individual BCDs for each channel.

The background-subtracted net source counts were then transformed
into physical units by multiplying by the flux conversion factor 
and then by the solid angle for each pixel, to yield fluxes for
each source. This flux density is then referenced against that of Vega 
in each IRAC
channel to put our photometry on a Vega-relative system.  Because we 
used a source aperture of 4 native IRAC pixels in radius, an aperture 
correction must be applied to the photometry.  Because the correction for
a 4 pixel radius source aperture is not given in the IRAC Data Handbook,
the aperture correction for the photometry was empirically determined 
by comparing our source aperture to the standard 10 pixel radius 
aperture using 15 relatively bright, unsaturated targets from our program.  
Thus, the final calibration of our data used is that of Reach et al. 
(2005), combined with our own aperture correction.  All of the relevant 
numbers used to calibrate our photometry 
are summarized in Table 2.  Table 3 summarizes the IRAC photometry
for all of our targets.
%\bigskip

\subsection{Error Analysis}

The photometry for each of our targets were measured using both
the co-registered, co-added, and cosmic ray cleaned mosaic images, and 
the individual BCDs in the dither pattern.  We used the latter data
to examine the repeatability of the photometry with a single AOR
in order to gain some insight into errors associated with photometry 
of the co-added data.  Figure 2 shows the standard deviation of the 
photometry from the individual BCDs for each target for each 
IRAC channel.  All of our targets were observed using HDR mode,
meaning both a short and long frametime exposure was taken at each 
position in the dither pattern.  In general, we favored the use of the 
long exposure (30-second frametime) BCDs over the short exposure 
(2-second frametime) BCDs in our HDR AORs, unless the target object 
was saturated in the long exposure frames.  In a few cases where 
there was good signal-to-noise in both the long and short exposure 
frames, we found good agreement (within a few percent) in the calibrated 
photometry from 
both frametimes.  We find that the mean fluxes from the individual measures
compare very well to those measured in the mosaic images.  In fact,
we used this comparison as one of our checks of photometric consistency
in detecting outliers in photometry of our target in the individual frames,
induced by bad pixels or cosmic ray hits.  

Overall, we find that the dispersions 
of the individual measures used to create the mosaics are larger than 
those estimated using photon statistics and basic detector characteristics 
(e.g., read noise, gain, etc.) alone.  For IRAC channels 1 and 2 these are 
5 and 3 times larger respectively, while for channels 3 and 4 they 
are essentially comparable, with the individual measure dispersions being 
only $\sim$20\% larger on average in each channel than the basic photon 
statistics.  This is probably due in part to 
some camera properties that have not been fully characterized in the 
current calibration.  These include intra-pixel phase sensitivity variations,
variations in the pixel solid angle due to geometric distortions introduced 
by the telescope and camera optics, and variations of the spectral 
response over the arrays due to the tilt of the filters with respect to 
the optical path and the relatively wide field of view of the camera.  For 
the latter two issues, the corrections are defined 
to be unity at the center of the arrays (Reach et al. 2005).  Thus, while 
all of these effects will act to increase the dispersion of the individual 
measurements in our dithered data, because our observing strategy for this 
program keeps the target object within $\sim$38 arcseconds of the center 
of the 
arrays, these additional uncertainties should be minimized.

We have chosen to use the standard deviation of the individual BCD measures 
for each source in each channel as the uncertainty of the photometry.
These are the uncertainties quoted in Table 3.  The number of BCDs used
to construct the co-added frame for each channel and to calculate the standard 
deviation are also listed for each source and channel in 
Table 3.  In terms of absolute calibration to the Vega-relative magnitude
system, the uncertainty of the IRAC photometry, 
convolved with the additional uncertainty introduced by our aperture 
correction, is about 3\% in IRAC channels 1$-$3 and 5\% in channel 4 
(Reach et al. 2005).

\section{IRAC Magnitudes and Colors for Late-M, L, and T dwarfs}

Molecular features found in the infrared for low mass stars 
and sub-stellar mass objects have strengths that are a strong function 
of temperature 
and pressure.  With the appropriate choice of photometric bandpasses, 
one can study trending as a function of color, magnitude, and spectral 
type as a part of characterizing the atmospheres of these objects.
For example, in the near-IR, the spectra of L dwarfs are characterized by
absorption from CO and H$_2$O, while the T dwarfs are dominated by 
broad absorption bands of CH$_4$ and H$_2$O, as well as collisionally induced 
absorption (CIA) by H$_2$.  Thus in near-IR photometry, M and L dwarfs become 
redder with decreasing \teff in $J-H$ and $H-K$.  The L to T dwarf transition
occurs as the silicate and iron condensates (clouds) become buried at 
increasing depth in the late-L dwarfs.  H$_2$O absorption 
begins to dominate the near-IR spectrum, leading to a bluing of the 
near-IR colors through the early-T types.  The colors 
then become even bluer from early-T to late-T with the onset and growth of 
CH$_4$ absorption and CIA H$_2$ in $K$.  The overall result is that the 
$J-H$ and $H-K$ colors for T dwarfs become bluer with increasing spectral 
subtype, becoming degenerate with the colors of higher mass K and M dwarfs.  

The IRAC filters were selected primarily to provide contiguous
bandpasses from $\sim$3 to 10 \mic for the determination of photometric 
red-shifts for extragalactic objects as part of the primary {\it Spitzer} 
mission 
objectives (Fazio et al. 2004).  However, these bandpasses were also defined,
in part, to 
provide diagnostics for the study of sub-stellar mass objects.  The bandpass
for IRAC channel 1 includes much of the CH$_4$ fundamental absorption band
($\sim$3.3 \mic).  Channel 2 includes the continuum peak that is 
present for all stars cooler than 3000 K, making this the most sensitive 
IRAC channel for the study of sub-stellar objects.  Channel 2 also contains
the broad but shallow CO fundamental absorption band ($\sim$4.7 \mic),
whose presence in the T dwarfs provides evidence for non-equilibrium
chemistry models (see discussion in section 5).  
Channel 3 includes H$_2$O absorption. 
Finally, for channel 4 the most important molecular absorption in this 
bandpass is due to CH$_4$.  

IRAC provides very precise fluxes in a wavelength regime that 
is poorly studied to date.  For this reason, we believe it is useful 
to examine the IRAC data from several perspectives.  First, we take a 
purely empirical viewpoint and limit ourselves to only the IRAC
photometry to explore the IRAC colors of M, L, and T dwarfs.  Next,
we combine IRAC and near-IR photometry to examine the location of 
M, L, and T dwarfs in both color-color and color-magnitude diagrams.
In section 4.3 we show the variations of near- and mid-IR colors with 
spectral type.  Finally, in section 5, we compare color-magnitude
diagrams for M, L, and T dwarfs with theoretical models.

\subsection{Color-Magnitude and Color-Color Diagrams}

\subsubsection{IRAC Photometry}

Star-formation studies often use plots of [3.6]-[4.5] versus [4.5]-[5.8] 
or [3.6]-[4.5] versus [5.8]-[8.0] in order to identify young stars with 
warm circumstellar dust disks and envelopes.  This technique is particularly 
effective because normal stars have essentially Rayleigh-Jeans spectra at 
IRAC wavelengths, and hence have colors very near zero.  Stars with 
warm dust disks have significantly red colors in IRAC wavelengths, and 
thus separate from older, dust-free stars.  Late-M, L, and T dwarfs, however, 
depart from this convenient scenario because they are cool enough 
to have photospheres sufficiently polluted by molecules that their spectra 
depart greatly from black-bodies.  

As shown in Figure 3, L dwarfs have 
[3.6]-[4.5] colors that remain close to zero, but have [4.5]-[5.8] colors 
that become significantly bluer.  For the T dwarfs, the strong redward 
trend in [3.6]-[4.5] 
is the result of CH$_4$ absorption removing increasing amounts of flux 
from the 3.6 \mic bandpass as \teff decreases.  On the other hand,
[4.5]-[5.8] trends strongly blueward for the T dwarfs, reflecting the 
strengthening of H$_2$O absorption with decreasing \teff in the 
5.8 \mic bandpass, relative to the CO absorption in the 
4.5 \mic bandpass.

For [3.6]-[4.5] versus [5.8]-[8.0] (Figure 4), the 
M and L dwarfs show about the same range of color in [5.8]-[8.0]
as in [4.5]-[5.8], but in this case there appear to be more overlap 
in the dispersion between these two types.  Again though, it is the T 
dwarfs that stand out in this plot, with both the 
[3.6]-[4.5] and [5.8]-[8.0] color indices trending strongly redward
for decreasing \teff.  For the latter color index, the H$_2$O 
feature in the 5.8 \mic bandpass apparently decreases the flux in this 
filter faster than the weaker CH$_4$ absorption band in the broader
IRAC 8.0 \mic bandpass.

To further explore the trending in the IRAC colors, we provide several 
variants on one specific IRAC color-color diagram - a plot of 
[4.5]-[5.8] versus [3.6]-[8.0].  In the first of these plots, 
Figure 5a shows our sample (excluding binaries and spectrally peculiar
dwarfs) with symbols corresponding to the object's
spectral type.  Figure 5a illustrates that the three spectral types inhabit 
relatively distinct parts of color-color space, with the [3.6]-[8.0] color 
becoming progressively redder for later spectral types (i.e., that it 
correlates with effective temperature), whereas the [4.5]-[5.8] color is 
approximately constant for M and L dwarfs, but turns sharply blueward with 
later T types.   There is not a one-to-one correspondence between spectral 
type and IRAC color however -- the M and L types overlap in their IRAC colors, 
whereas the T dwarfs appear to have a very large scatter in their 
colors.   Are these effects simply ``noise'', or are they indicative of 
complexity in the spectra of these stars beyond that attributable to just 
spectral type?

To address these questions, we construct several other versions of 
Figure 5a but now isolating specific spectral subtype ranges.  Figure 5b 
shows the location of ``early'' and late M dwarfs in this color-color plane, 
and illustrates that at least for our sample of stars, field M dwarfs with 
spectral type M8 and earlier have IRAC colors essentially consistent with 
zero, with only the M8.5 and later M dwarfs having significantly non-zero 
IRAC colors.  It is only the stars in the latter sub-type range that 
overlap in color with the L dwarfs.   Figure 5c similarly isolates several 
L dwarf subtypes.   We have combined the L0 through L3.5 dwarfs into one 
group because we see no significant variation in IRAC color within this 
subtype range; however, the IRAC colors of these early L dwarfs are 
significantly different from the colors of M dwarfs with type earlier
than M8.  The L4 and L4.5 dwarfs appear to be a transition class, with 
two dwarfs of this type having IRAC colors like those or earlier type L 
dwarfs, while the other two are significantly redder, particularly in 
their [4.5]-[5.8] color.   The L7.5 to L8 dwarfs have redder colors 
than earlier L dwarfs, and have [4.5]-[5.8] colors that are redder, 
on average, than even the early T dwarfs.   

Figure 5d highlights the IRAC colors of T dwarfs.   The early T dwarfs 
have [3.6]-[8.0] colors that overlap with the latest L dwarfs, but 
they are displaced blueward in [4.5]-[5.8] color from those dwarfs.  
In general, the T dwarfs show a progression, with their [3.6]-[8.0] 
color becoming redder and their [4.5]-[5.8] color becoming bluer with 
later T subtype.  The T5 and T6 subtypes appear to be transitional, with 
a relatively large spread in [3.6]-[8.0] color for the T5 dwarfs and,
in comparison, colors for the two T6 dwarfs in our sample that are closer
to those of the T4 dwarfs than the T7$-$T8 dwarfs.  

Finally, in Figure 5e, we show the relative colors of those objects
in our sample that are considered spectrally peculiar or unusual.  
These include 
the lone sdL in our sample (2MA0532+8246) and the lone T dwarf with a 
peculiar designation in its spectral type (T6p, 2MA0937+2931).  The other 
T dwarfs in the figure have been noted in the literature as being spectrally 
unusual.  For the latter, how much of their peculiarity is due to 
atmospheric anomalies (i.e., metallicity) or multiplicity remains to be seen.  
For example, only recently has the unusual spectrum of 2MA0423-0414 been
revealed to be a composite spectrum of an $\sim$L6 plus $\sim$T2 close 
binary system (Burgasser et al. 2005).  Nevertheless, these objects as a 
whole seem to have discrepant colors when compared to other L and T dwarfs, 
particularly for 2MA0532+8246 (sdL), 2MA0937+2931 (T6p), and
2MA1217-0311 (T7.5).

\subsubsection{IRAC Photometry Combined with $JHK$ Photometry}

The combination of IRAC broadband photometry with near-IR 
photometry\footnote{For this comparison to work, it is necessary 
for all of the 
near-IR $JHK$ photometry to be on the same photometric system.
While a number of $JHK$ systems are reasonably similar, with only
a few percent difference in the fluxes for observations of normal stars
(e.g., Leggett et al. 2002, Bessell \& Brett 1988),
these differences are magnified for low mass and sub-stellar mass
objects, where 
strong molecular absorption bands in the infrared lead to 
$JHK$ magnitudes that are highly dependent on the bandpasses of 
the filters used.  The recent introduction of the MKO near-IR photometric 
system (Tokunaga et al. 2002) has made this problem especially 
acute for the L and T dwarfs, where differences of $\sim$20\% or more are 
observed between MKO photometry and that of other $JHK$ systems
for the same object.  
For all figures in this paper that use $JHK$ photometry, we 
have transformed these data to the 2MASS system using relations from 
Stephens \& Leggett (2004).  While we recognize that the MKO 
system has been endorsed by the IAU as the preferred photometric 
system for ground-based near-IR observations, the vast majority of 
the $JHK$ data 
for the objects in our sample are already in the 2MASS system.}
breaks the degeneracy of the M and L dwarfs colors.
In a plot of $K_S$-[3.6] versus [3.6]-[4.5] (Figure 6), the 
$K_S$-[3.6] color index shows a clear trend redward for the 
M and L dwarfs, well correlated with spectral type.  For the T dwarfs,
this same index trends over a more narrow range of color for 
increasing subtype.  Although the $K_S$-[3.6] color index is almost
degenerate, the [3.6]-[4.5] color index still spreads them nicely 
in Figure 6.  Because pressure induced H$_2$ absorption affects the 
$K_S$ band in particular and is stronger if gravity is high or 
metallicity is low, this can introduce scatter in $K_S$.  This is
likely the major contributor to the large scatter in color seen
with the same spectral subclass in the mid- and late-T dwarfs
(see also section 4.3 for the trending of color versus spectral
type for these two color indices).  For $K_S$-[4.5] versus [3.6]-[4.5] 
(Figure 7), the same breaking of degeneracy and range of color is seen 
for the M and L dwarfs in $K_S$-[4.5] as for $K_S$-[3.6]; however, in this 
case the T dwarfs show a strong redward trend, with a spread of $\sim$2.5 
magnitudes in $K_S$-[4.5].  The relative amount of scatter in the 
color index for the mid- and late-T dwarfs remains the same as 
that seen for $K_S$-[3.6] in Figure 6.

\subsection{Absolute Magnitude versus Color}

For those objects in our program with trigonometric parallax
measurements, we present two color-magnitude plots
that could be potentially useful for identifying low-mass 
dwarfs in the field or as companions to other stars, using 
a combination of IRAC and near-IR photometry (some additional 
IRAC color-magnitude diagrams focusing on the T dwarfs are also 
presented in section 5).

In Figure 8, M$_{K_S}$ versus $K_S$-[4.5] shows that the colors of M, L, 
and T dwarfs follow a relatively smooth progression with 
increasing type.  This is similar to the trending seen for 
M$_{K_S}$ versus $K_S$-$M'$ as reported by Golimowski et al. 2004a.
On the other hand, in Figure 9, M$_{K_S}$ versus $J$-[4.5]
shows almost 2.5 magnitudes of range for the M and L 
dwarfs, while for the T dwarfs the $J$-[4.5] color is
essentially degenerate with the L dwarfs (and with themselves,
the early-T dwarfs falling in the transition area between the 
L dwarf and T dwarf plateaus, being degenerate not only 
with the mid-L dwarfs but also with mid-T types).  It is only
the intrinsic difference in luminosity that lifts the 
degeneracy for the $J$-[4.5] color index.

\subsection{Color versus Spectral Type}

In the same spirit as Leggett et al. (2002), Knapp et al. (2004), 
and Golimowski et al. (2004a), we summarize the trending of 
various color indices against spectral type (Figure 10) as
a series of ``postage stamp'' plots.  While most of this trending
has already been detailed earlier in this section in the color-color
and color-magnitude diagrams, plotting the indices against 
spectral type reveals some common themes.  In particular, most
of the plots in Figure 10 show breaks in the trending near
the L/T boundary.  The most dramatic example is in [3.6]-[4.5]
where the slow blueward trend through the M and early-L types
turns redward at mid-L (given the appearance of CH$_4$ absorption in 
the IRAC 3.6 \mic bandpass) and heads strongly redward through the
T types.  On the other hand, the $K_S$-[4.5] and [4.5]-[5.8] plots show 
breaks at the mid-T types.  For [4.5]-[5.8], the blueward turn at mid-T 
presumably happens due to the strengthening of H$_2$O absorption with 
later spectral type in the IRAC 5.8 \mic filter, while for $K_S$-[4.5], 
the onset of both CH$_4$ absorption (2.2 \mic overtone feature) and CIA 
H$_2$ in the $K_S$ band drives the color index redward for later types.
The redward trend of the $K_S$-[3.6] color index turns blueward as the 
fundamental CH$_4$ absorption feature at 3.2 \mic switches on at mid-L,
and then turns redward again at mid-T when this feature saturates and
2.2 \mic CH$_4$ absorption and CIA H$_2$ in the $K_S$ band assert
themselves at mid-T.

These color versus spectral type plots serve to show that 
in the previously presented color-color diagrams based on
IRAC photometry alone, much of the scatter in the T dwarf
colors come from the [4.5]-[5.8] and [5.8]-[8.0] color
indices.  The trending of color versus spectral type for 
the [3.6]-[4.5] color index is very smooth.  
While some of the scatter in the [5.8]-[8.0] color index could be
due to the interplay of H$_2$O absorption (starting in the $\sim$late-M
types) versus the 7.7 \mic overtone band of CH$_4$ in the 
IRAC 8.0 \mic filter (beginning at $\sim$mid-L types), it is also 
true that our fixed integration time AORs yield the highest 
quality data for IRAC channels 1 and 2, the most sensitive of the four 
IRAC channels, and thus systematically lower signal-to-noise data 
for our sample in IRAC channels 3 and 4 (i.e., some of the scatter 
seen in [5.8]-[8.0] might be due to the larger errors
on this color index compared to the others presented in Figure 10).

\section{Discussion}

As shown in the previous sections, the infrared colors of low mass stars 
and brown dwarfs are generally well-correlated with their spectral types.  
In turn, 
the spectral types of M, L, and T dwarfs are reasonably well-correlated 
with their effective temperatures, though to a lesser degree than for 
higher mass objects because molecular
chemistry and particulate clouds compete aggressively in determining the
detailed spectral energy distributions of cool dwarfs.  Theoretical models
are beginning to be able to match the observed spectral properties of very
cool dwarfs, and we have used those models to help interpret the IRAC 
photometry in Section 4.

While a detailed theoretical analysis of these new {\it Spitzer}/IRAC data is 
beyond the scope of this paper, there are a few conclusions 
of a general nature that can be extracted at this preliminary stage.   
To do so, we have generated three color-magnitude plots (Figures 11--13) 
for most of the T dwarfs in our sample with parallaxes and superposed 
synthetic magnitudes and colors derived from theoretical spectra for an 
effective temperature (\teff)
range from 700 K to 1300 K in steps of 100 K at three gravities.  
These gravities are 10$^{4.5}$ cm s$^{-2}$ (green circles), 10$^5$ cm s$^{-2}$ (red triangles), 
and 10$^{5.5}$ cm s$^{-2}$ (blue squares).  Figure 11 depicts the absolute magnitude 
in the IRAC 3.6 \mic bandpass versus the [3.6]-[4.5] color.  The legend on 
this 
figure indicates the T dwarfs included on this figure and they are numbered 
in order of spectral sub-type.

As Figure 11 demonstrates, despite the fact that the numbering is in order 
of increasing spectral sub-type, the 3.6 \mic IRAC data are not monotonic 
in either magnitude or color.  That they are not in order in magnitude may 
in part be ascribed to multiplicity.  However, this cannot be the 
explanation for the majority of these T dwarfs.  Furthermore, the [3.6]-[4.5] 
colors too are non-monotonic in spectral type.  This is even more 
obvious in Figure 12, which portrays M$_{5.8}$ versus [5.8]-[8.0].  
For example, 
SDS1346-0031 and 2MA0727+1710 only differ by half a spectral sub-type, but 
they have very different colors.  

The non-monotonicity of these {\it Spitzer}/IRAC data is the clearest indication yet
that the T dwarfs are not a one-parameter family in \teff, but that more than
one parameter is influential in determining the spectroscopic type. 
Because the metallicity can not generically vary enough in the solar 
neighborhood to explain this, the extra parameter may be gravity.  This was 
the conclusion of Burrows et al. (2002) and Knapp et al. (2004) using 
different datasets at shorter wavelengths, and is confirmed here.  
For the T dwarfs, gravity will translate into mass (Burrows et al. 1997).  
From the 
comparison of the spread in the data in Figures 11 and 12 
with the spread with gravity of the theoretical models, we conclude that
a range of gravities of approximately a factor of five is represented 
in the extant T dwarf sample.  This can be converted into a range of 
masses from $\sim$70 M$_{Jupiter}$ to $\sim$15 M$_{Jupiter}$.  
The less massive objects would 
also be younger and a range of ages of about a factor of ten (0.2$-$0.3
to 10 Gyr) is 
indicated (Burrows et al. 1997; Burrows et al 2001).  The latter conclusion 
might be problematic, 
but such is indicated by our preliminary analysis.

Figure 13 is an HR diagram of IRAC M$_{4.5}$ versus
the IRAC [4.5]-[5.8] color.  The non-monotonicity of the data seen in
previous plots 
color survives in this plot as well.  However, as is abundantly clear
in the figure, the theoretical models do not fit these data.  Because CO 
has a strong spectral feature at $\sim$4.67 \mic,
one can interpret the discrepancy between theory and the IRAC data as
an indication that equilibrium CH$_4$/CO chemistry
underestimates the abundance of CO in T dwarf atmospheres.
This conclusion was already reached by Golimowski et al. (2004a) 
using $M$-band measurements from the ground.   The magnitude of the 
discrepancy translates into an overestimate
in the $\sim$4--5 \mic flux by factors of 1.5 to 3.0.

It should be noted that the opacities of water in the mid-infrared
are still being studied and refined (Schwenke 2002).  
Continuing ambiguity in the H$_2$O opacities are a
source of systematic theoretical error, with the product
of the O abundance and the H$_2$O opacity directly connected to
the goodness of fit at 5.8 and 8.0 \mic for the later T dwarfs.
Moreover, at the lowest \teff represented here, the condensation
of K into KCl and Na into Na$_2$S could introduce hazes with interesting
optical depths.  Finally, the early T dwarf spectra show signs of 
silicate clouds (Stephens, Marley, \& Noll 2001; Marley et al. 2002; Burrows 
et al. 2002) not included in the models we have presented.  Hence, 
the study of sub-stellar mass object atmospheres is still in its 
formative years.

Nevertheless, as Figure 14 implies, one can be encouraged
that the basics are falling into place. Figure 14  
compares a theoretical T dwarf spectral model
with \teff/g = 750 K/10$^{5}$ cm s$^{-2}$ (Burrows et al. 2002; 
Geballe et al. 2001) with the four IRAC fluxes measured
for the T7.5 dwarf GJ 570D. This model was generated in 2002 for GJ 570D to
fit its optical spectrum {\it shortward} of 1.0 \mic
(Burrows et al. 2002).  A slight ($\sim$40\%) discrepancy in the 4.5-\mic 
bandpass,
attributable to a CO abundance excess in its atmosphere, is visible.
Despite this, Figure 14, and Figure 7 in the related paper by
Burrows et al. (2002), together represent an acceptable fit
from 0.6 \mic to $\sim$8.0 \mic and indicate
that the IRAC data were successfully anticipated.

\acknowledgements
This work is based (in part) on observations made with the {\it Spitzer Space
Telescope}, which is operated by the Jet Propulsion Laboratory, California
Institute of Technology under NASA contract 1407. Support for the IRAC 
instrument was provided by NASA under contract number 1256790 issued by JPL.
We thank Sandy Leggett (our referee) and Adam Burgasser
for providing us with insightful and helpful
comments on our manuscript.
This publication makes use
of data from the Peters Automated Infrared Imaging Telescope (PAIRITEL),
which is operated
by the Smithsonian Astrophysical Observatory (SAO) and was made possible by
a grant from the Harvard University Milton Fund, the camera loan from
the University of Virginia, and the continued support of the SAO and 
the University of California, Berkeley.
AB wishes to acknowledge NASA for its financial support via grant
NNG04GL22G. Furthermore, AB acknowledges support through the Cooperative 
Agreement \#{NNA04CC07A} between the University of Arizona/NOAO LAPLACE node 
and NASA's Astrobiology Institute.  This research has made use of the L and T 
dwarf compendium housed at DwarfArchives.org and maintained by 
Chris Gelino, Davy Kirkpatrick, and Adam Burgasser.  This research also 
has made use of the SIMBAD database, operated at CDS, Strasbourg, France
and data from the Two Micron All Sky Survey, a joint project of the 
University of Massachusetts and the Infrared Processing and Analysis
Center.
SMS acknowledges support for this research by the SAO Summer Intern 
REU Program, funded by the National Science Foundation and the Smithsonian 
Institution.

%\clearpage

%%
%% Figures
%%
\clearpage
\centerline{\ .}
\bigskip
\bigskip
\bigskip
\begin{figure}[h]
\figurenum{1}
%\plotone{./f1.eps}
\includegraphics[angle=270,scale=0.65]{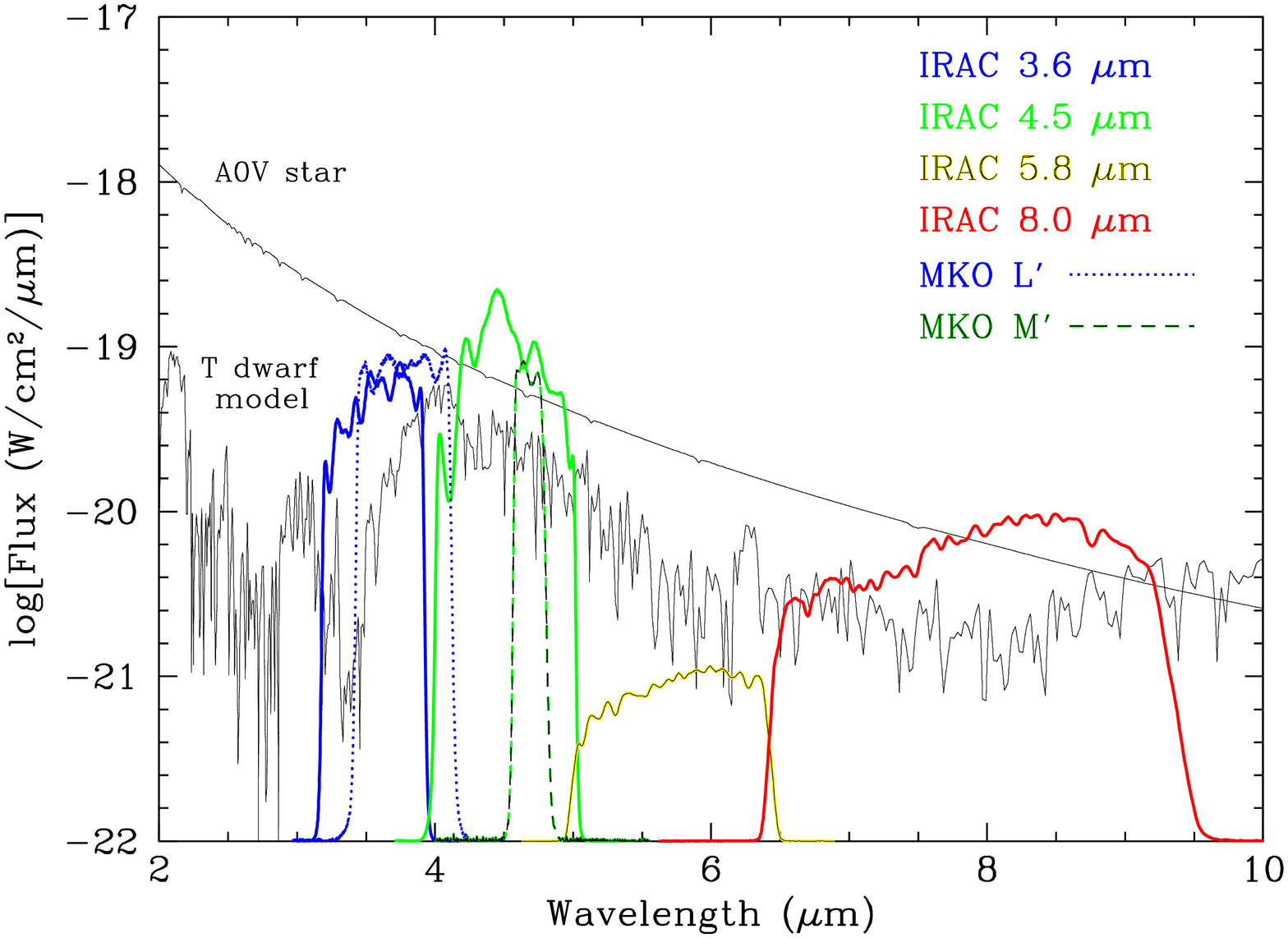}
%\centerline{Fig. 1. ---}
\caption{A comparison of the IRAC bandpasses with the MKO $L'$
and $M'$ filters in recent use for ground-based studies of 
M, L, and T dwarfs.  Also shown for comparison is a model
spectrum for a \teff = 950K, $g=10^5$ cm s$^{-2}$ $\sim$mid-T dwarf
(Burrows et al. 2002) and that for an A0 V star.  The IRAC 
bandpasses have been scaled to show their relative sensitivity.}
\end{figure}

\clearpage
\centerline{\ .}
\bigskip
\bigskip
\bigskip

\begin{figure}[h]
\figurenum{2}
\plotone{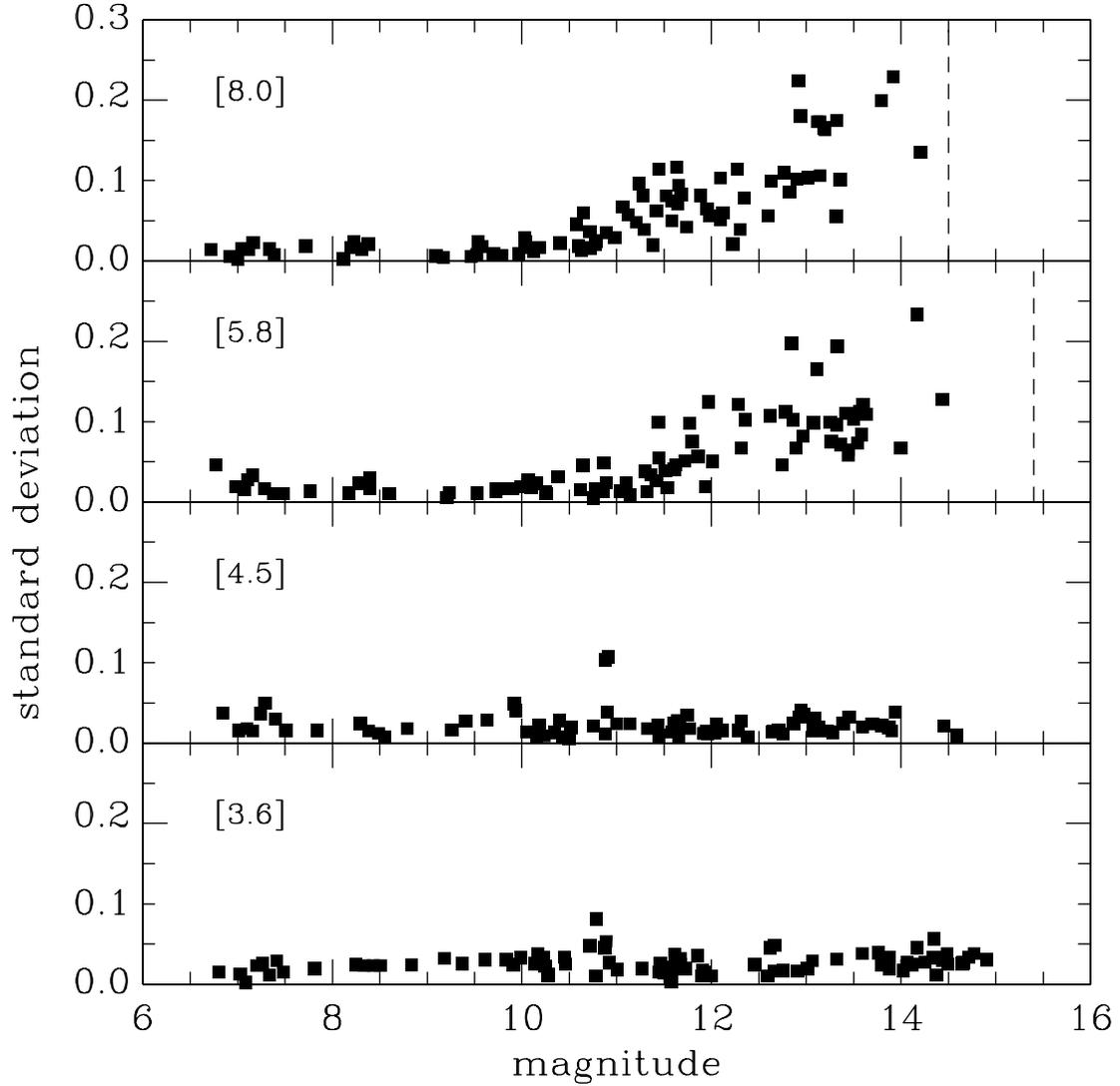}
\caption{Repeatability of photometry for targets versus their 
calibrated 
magnitudes for the four IRAC channels.  Shown here are the standard 
deviations of the individual observations in the AOR dither pattern
(five for most objects) versus the calibrated magnitude derived from 
the co-addition of those individual observations.  For the 5.8 and 
8.0 $\mu$m channels, the 3$\sigma$ limiting magnitude for a high zodiacal 
background for a 30-second frametime is indicated (15.4 and 14.5 respectively)
(Additional details can be found in the IRAC Data Handbook at 
http://ssc.spitzer.caltech.edu/irac/data.html).  For the 3.6 and 4.5 
$\mu$m channels, the 3$\sigma$ limiting magnitudes would be located 
off the right of the plots at 18.9 and 17.9 respectively.}
\end{figure}

\clearpage
\centerline{\ .}
\bigskip
\bigskip
\bigskip

\begin{figure}[h]
\figurenum{3}
%\plotone{./f3.eps}
\includegraphics[angle=270,scale=0.65]{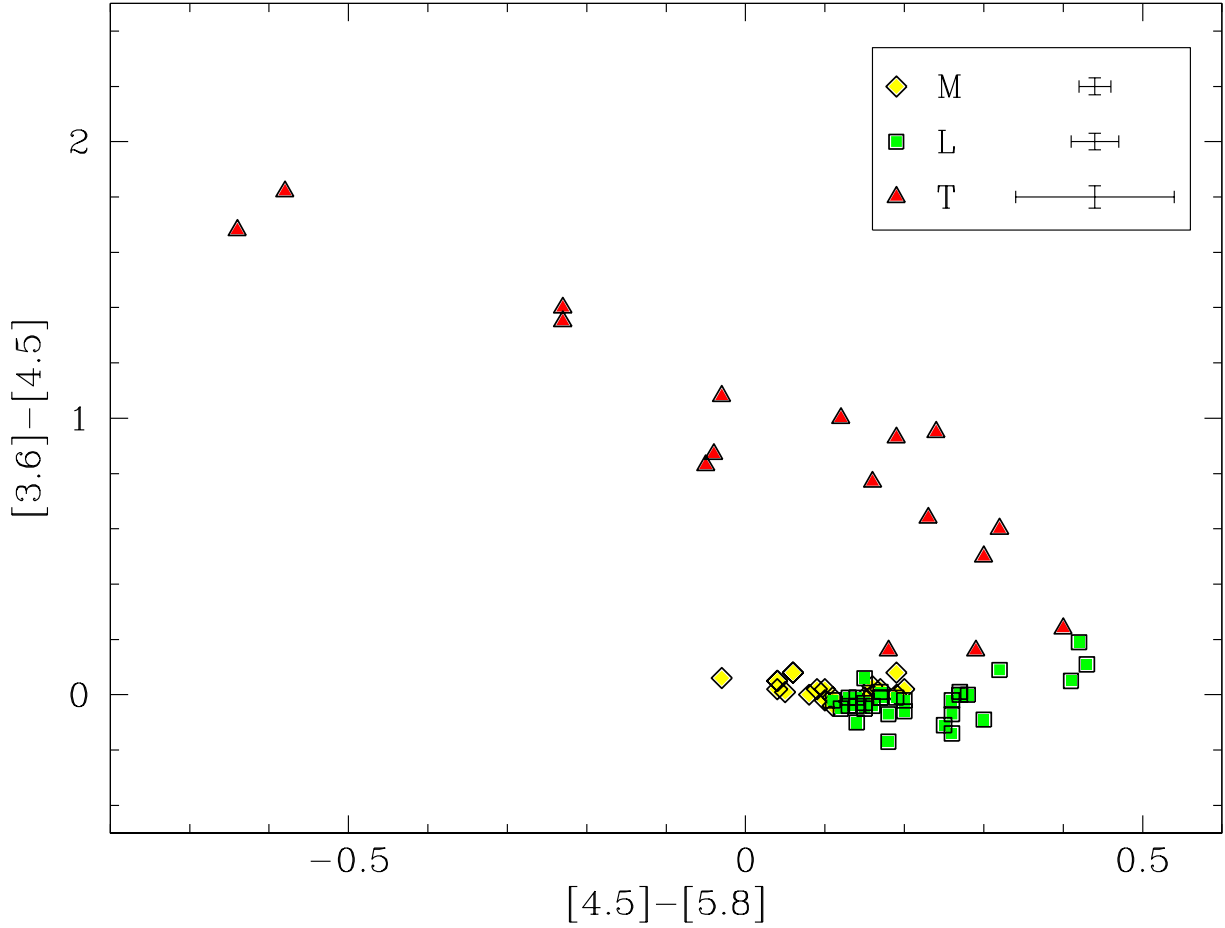}
\caption{[3.6]-[4.5] versus [4.5]-[5.8] color-color diagram for
all of the sources in our M, L, and T dwarf sample, excluding known 
binaries and spectrally peculiar objects.  As indicated in the legend, 
different plot symbols and colors are used to represent the M dwarfs 
({\it yellow diamonds}), L dwarfs ({\it green squares}), and 
T dwarfs ({\it red triangles}).  The error bars to the right of each 
plot symbol in the legend represent the median values for each spectral
type for the color indices used in this figure.}
\end{figure}

\clearpage
\centerline{\ .}
\bigskip
\bigskip
\bigskip

\begin{figure}[h]
\figurenum{4}
%\plotone{./f4.eps}
\includegraphics[angle=270,scale=0.65]{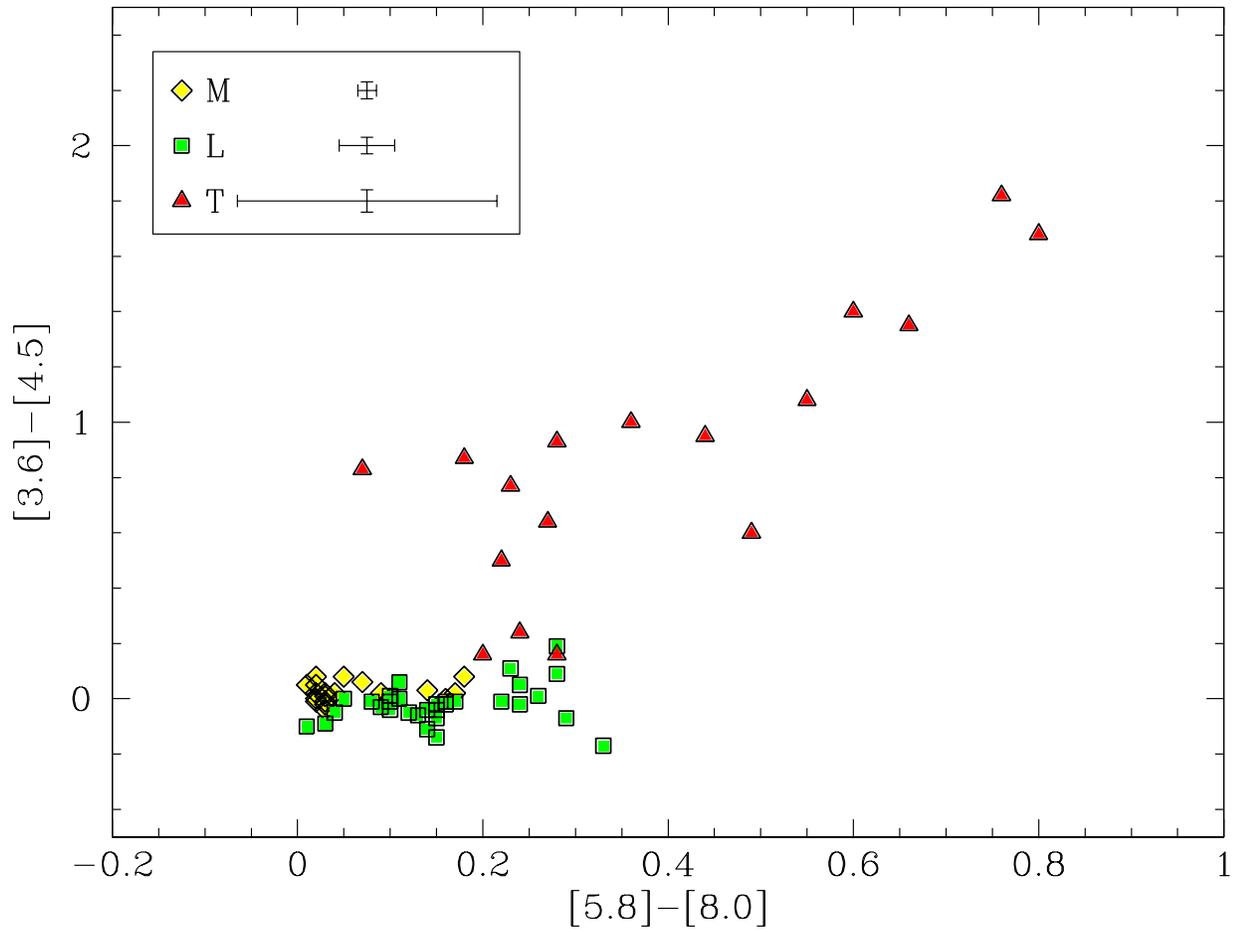}
\caption{Similar to Figure 3 except in this case we show
[3.6]-[4.5] versus [5.8]-[8.0] color-color diagram for the same 
objects.}
\end{figure}

\clearpage
\centerline{\ .}
\bigskip
\bigskip
\bigskip

\begin{figure}[h]
\figurenum{5a}
%\plotone{./f5a.eps}
\includegraphics[angle=270,scale=0.65]{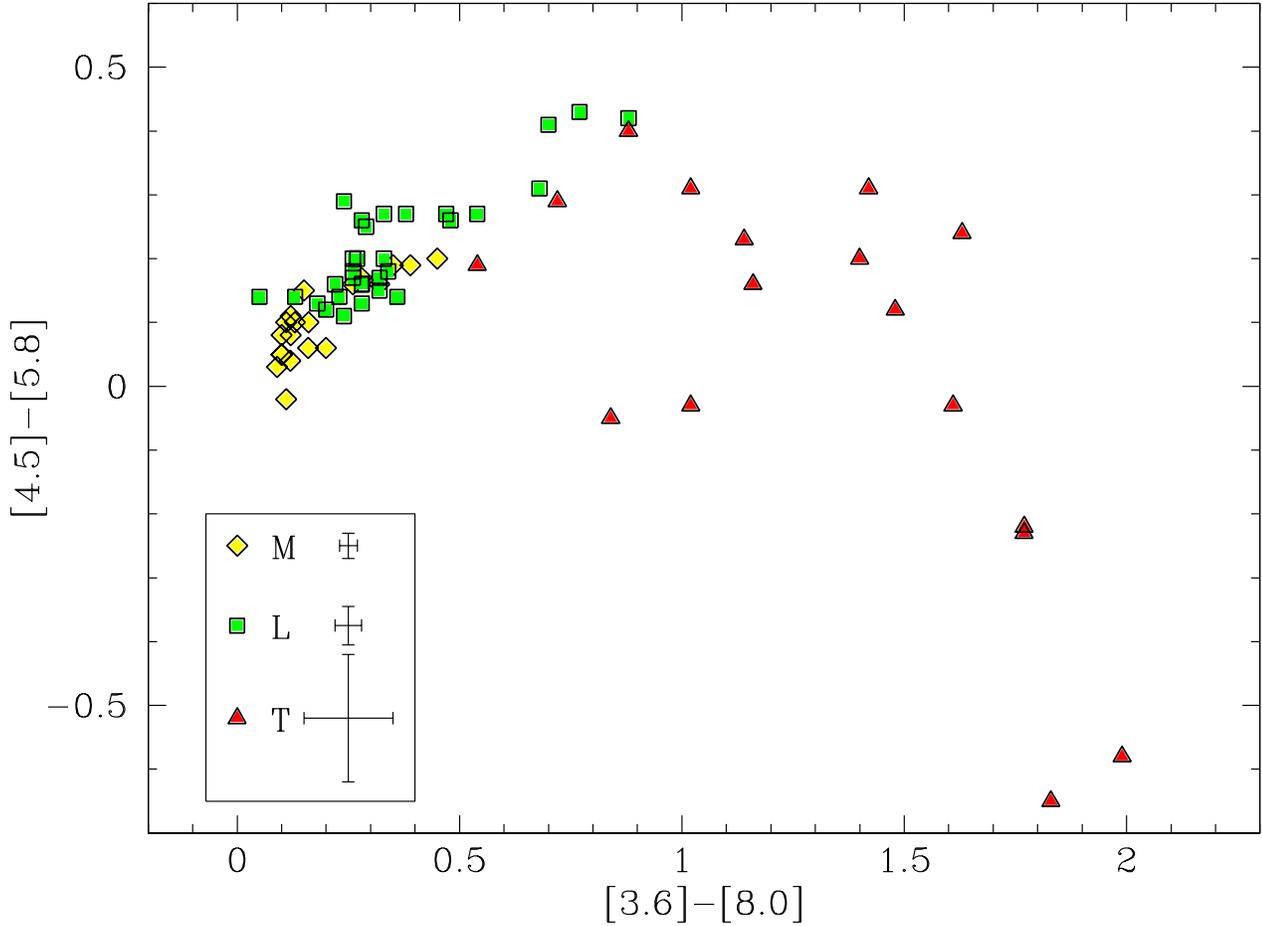}
\caption{[3.6]-[8.0] versus [4.5]-[5.8] color-color diagram.
{\bf (a)} Same plot symbols and objects as plotted in Figure 3.  
{\bf (b)} Same data as in (a) except colored symbols 
show region of color-color space occupied by spectral types $<$M8 
({\it yellow squares}) and M8$-$M9 ({\it green circles}).
{\bf (c)}  In this case the colored symbols show a clear separation
in color-color space for early-L types (L0$-$L3) ({\it blue squares})
and late-L (L5$-$L8) ({\it red triangles}).  The L4 dwarfs ({\it cyan
circles}) appear to
form a transition class between the early and late types.
{\bf (d)}  The break out of the T dwarfs shows that for the 
early types T0$-$T1 ({\it blue squares}), T2$-$T3 ({\it cyan triangles}),
and T4 ({\it crosses}), there is a trend towards redder [3.6]-[8.0]
color with increasing spectral type, but little discernible change 
in the [4.5]-[5.8] color over these same types.  The late-T 
types T7$-$T8 ({\it red asterisks}) are much redder in [3.6]-[8.0]
and bluer in [4.5]$-$[5.8] than the early types. 
The T5 ({\it green circles}) and T6 ({\it yellow diamonds}) subtypes appear 
to be transitional, with a relatively large spread in [3.6]-[8.0] color 
for the T5 dwarfs and redder [4.5]-[5.8] colors for the T6 dwarfs.
{\bf (e)}  Spectrally peculiar objects in our sample.  Each object
is labeled with its spectral type and errors bars representing the 
uncertainty of the photometry.  The colored symbols are data for the same 
objects as shown in (a).}
\end{figure}

\clearpage
\centerline{\ .}
\bigskip
\bigskip
\bigskip

\begin{figure}[h]
\figurenum{5b}
%\plotone{./f5b.eps}
\includegraphics[angle=270,scale=0.65]{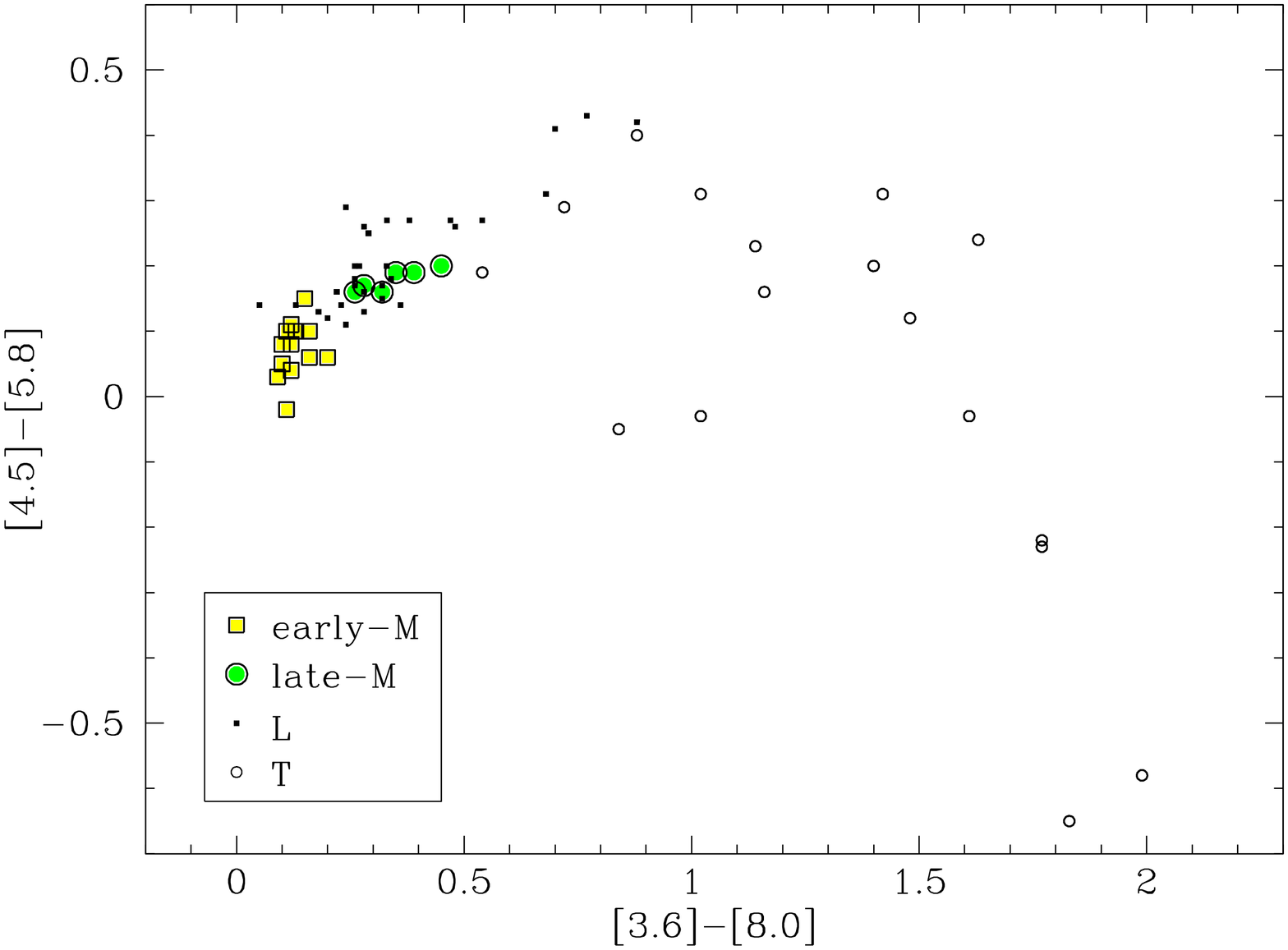}
\caption{\ }
\end{figure}

\clearpage
\centerline{\ .}
\bigskip
\bigskip
\bigskip

\begin{figure}[h]
\figurenum{5c}
%\plotone{./f5c.eps}
\includegraphics[angle=270,scale=0.65]{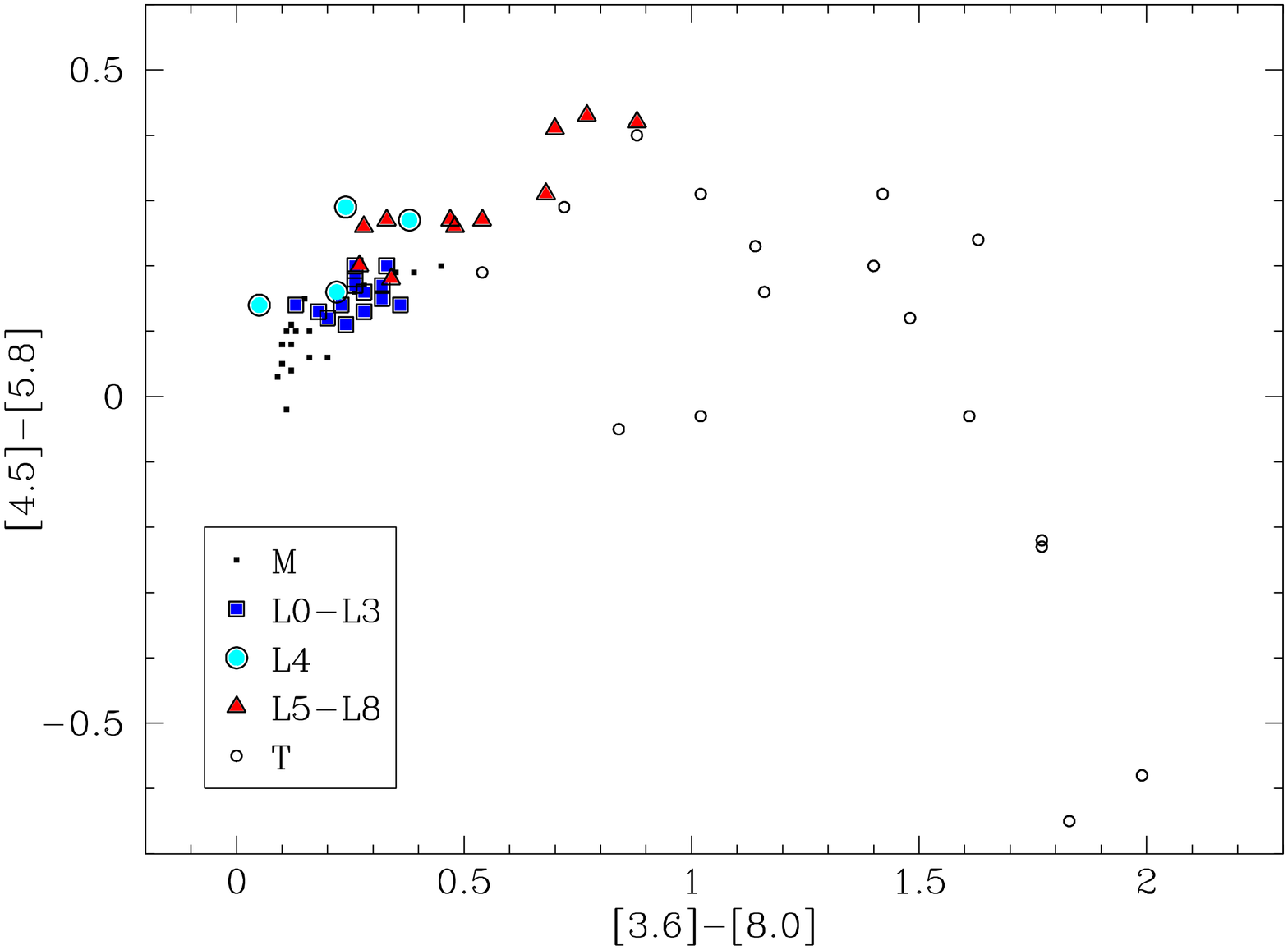}
\caption{\ }
\end{figure}

\clearpage
\centerline{\ .}
\bigskip
\bigskip
\bigskip

\begin{figure}[h]
\figurenum{5d}
%\plotone{./f5d.eps}
\includegraphics[angle=270,scale=0.65]{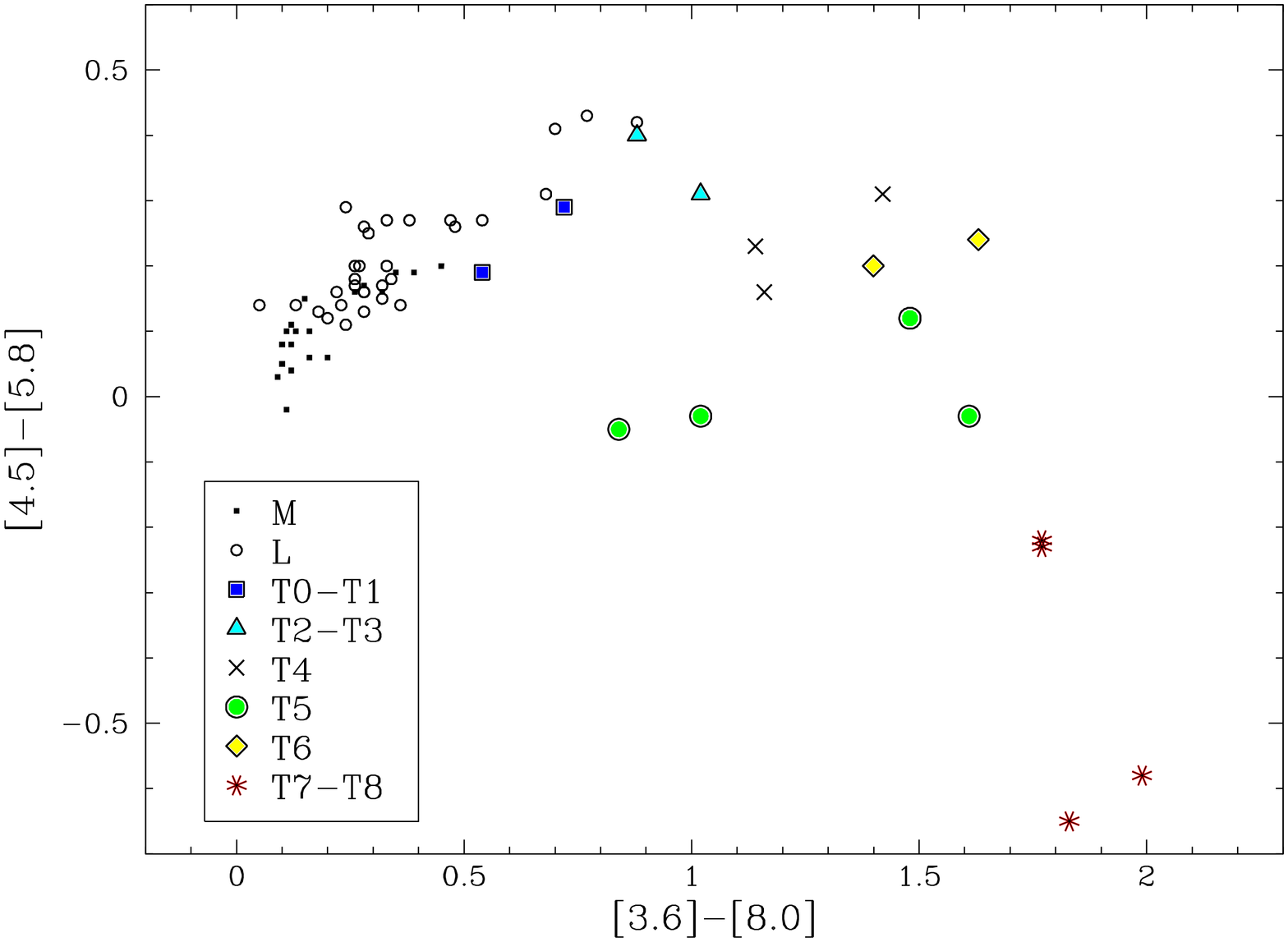}
\caption{\ }
\end{figure}

\clearpage
\centerline{\ .}
\bigskip
\bigskip
\bigskip

\begin{figure}[h]
\figurenum{5e}
%\plotone{./f5e.eps}
\includegraphics[angle=270,scale=0.65]{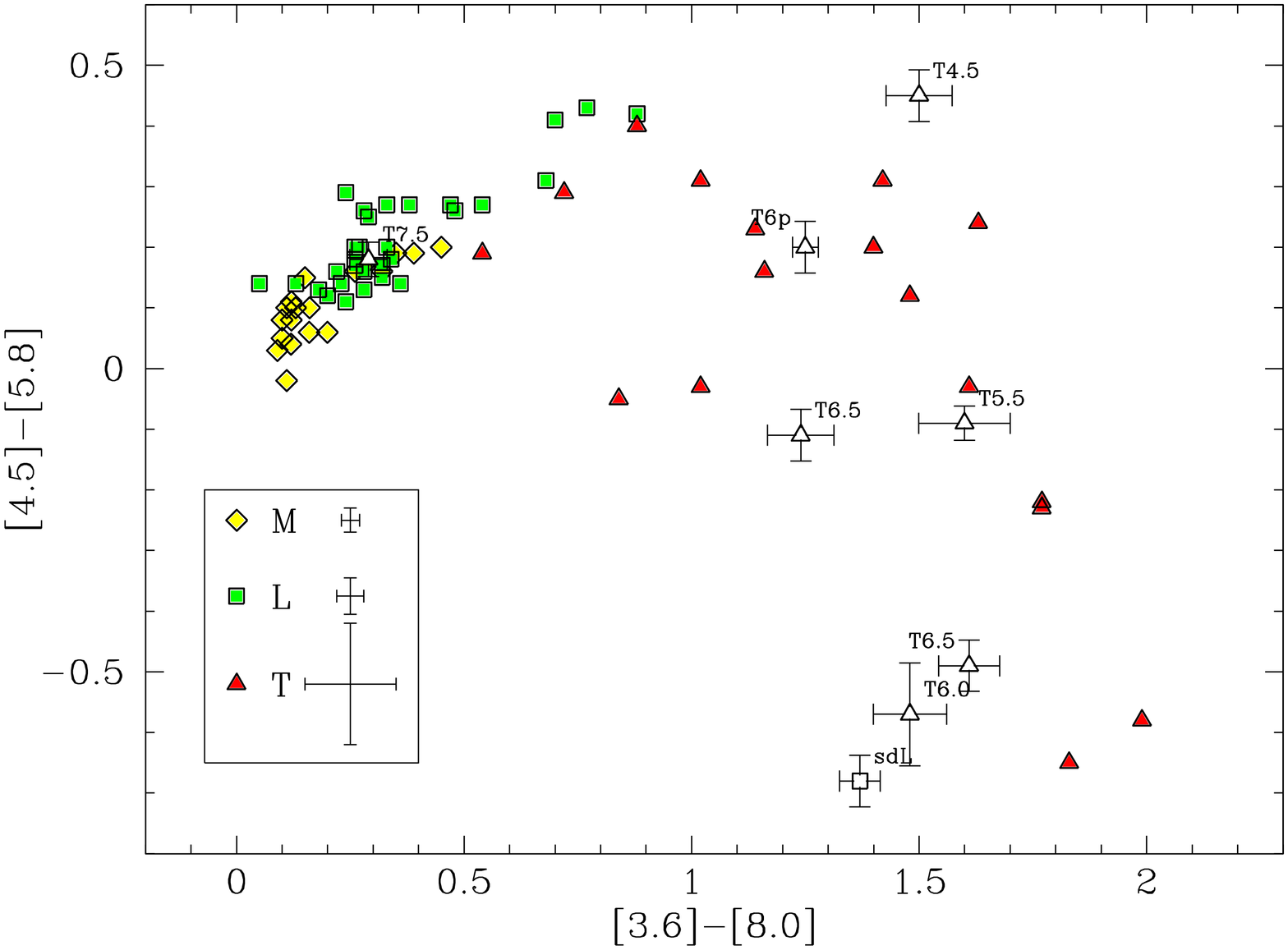}
\caption{\ }
\end{figure}

\clearpage
\centerline{\ .}
\bigskip
\bigskip
\bigskip

\begin{figure}[h]
\figurenum{6}
%\plotone{./f6.eps}
\includegraphics[angle=270,scale=0.65]{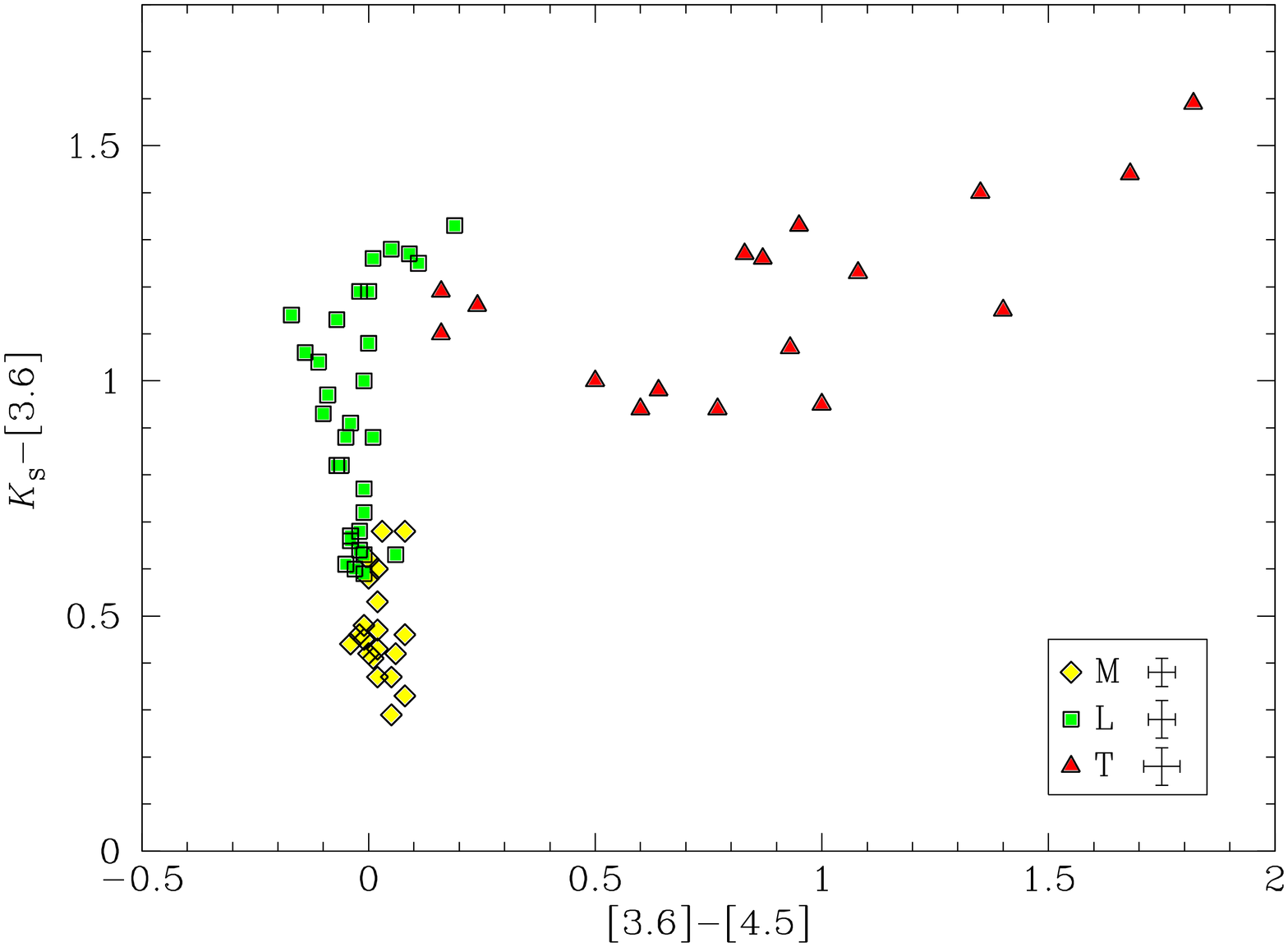}
\caption{$K_S$-[3.6] versus [3.6]-[4.5] color-color diagram for
all of the sources in our M, L, and T dwarf sample with $K$-band 
photometry, excluding known binaries and spectrally peculiar objects.  
All near-IR photometry has been converted to the 2MASS $JHK$ system
using the relations of Stephens \& Leggett (2005) (see additional 
discussion in the text).  The plot symbols are the same as those used in 
Figure 3.}
\end{figure}

\clearpage
\centerline{\ .}
\bigskip
\bigskip
\bigskip

\begin{figure}[h]
\figurenum{7}
%\plotone{./f7.eps}
\includegraphics[angle=270,scale=0.65]{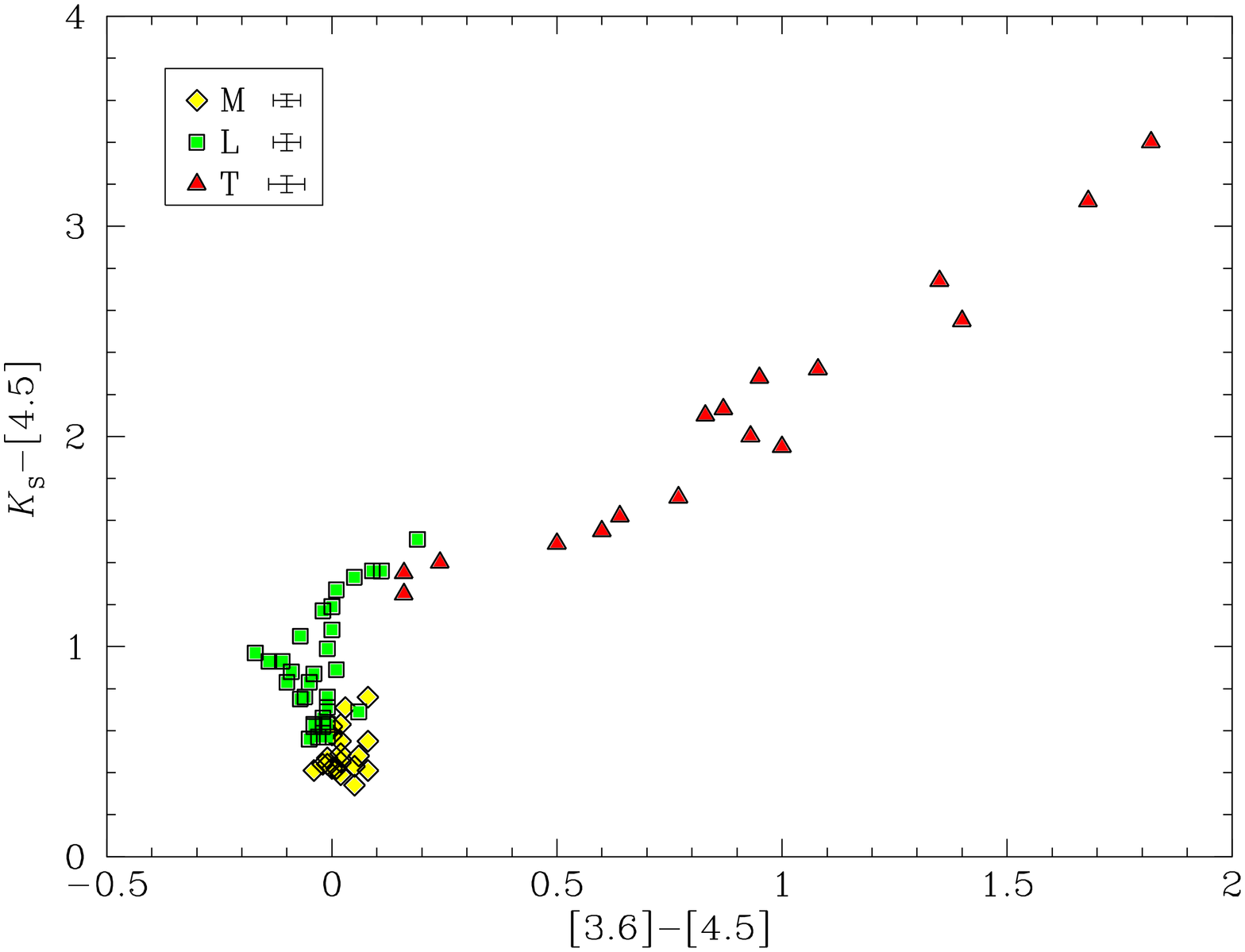}
\caption{Similar to Figure 6 except in this case we show
$K_S$-[4.5] versus [3.6]-[4.5] color-color diagram for the same objects.}
\end{figure}

\clearpage
\centerline{\ .}
\bigskip
\bigskip
\bigskip

\begin{figure}[h]
\figurenum{8}
%\plotone{./f8.eps}
\includegraphics[angle=270,scale=0.65]{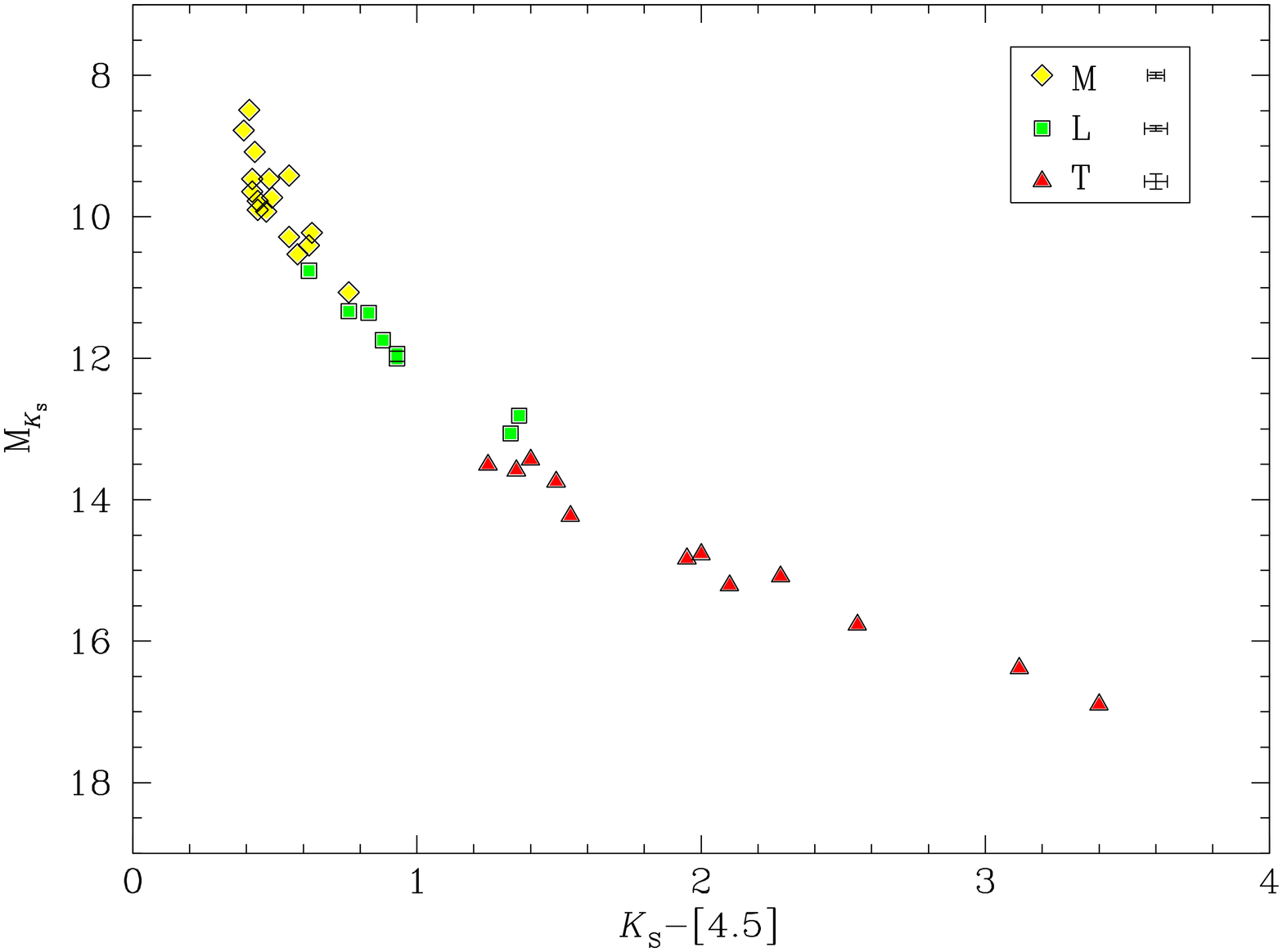}
\caption{M$_{K_S}$ versus $K$-[4.5] color-magnitude diagram for
all of the sources in our M, L, and T dwarf sample with $K$-band 
photometry and trigonometric parallax, excluding known binaries and 
spectrally peculiar objects.  As in Figure 6, all near-IR photometry 
has been converted to the 2MASS $JHK$ system.  Parallax references
can be found in Table 1. The plot symbols are the same as those used in 
Figure 3.}
\end{figure}

\clearpage
\centerline{\ .}
\bigskip
\bigskip
\bigskip

\begin{figure}[h]
\figurenum{9}
%\plotone{./f9.eps}
\includegraphics[angle=270,scale=0.65]{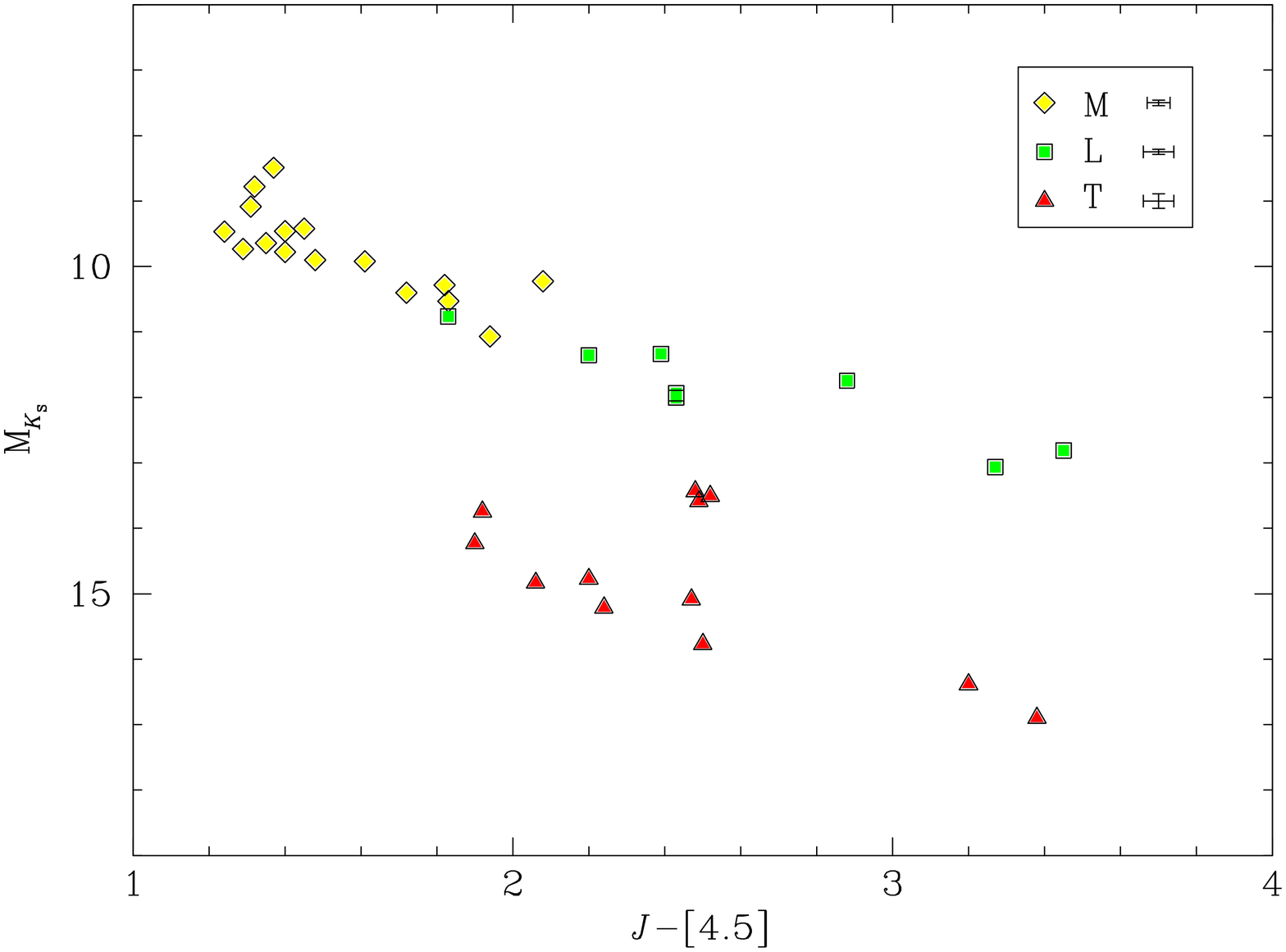}
\caption{Similar to Figure 8 except in this case we show
M$_{K_S}$ versus $J$-[4.5] color-magnitude diagram for the same objects.
The three T dwarfs in the region between the locus of the L dwarfs
and the parallel but fainter locus of T dwarfs below have spectral
types of T0.5, T1.0, and T2.0.
The sequence of T dwarfs below run from {\it left to right}
from T3.5 to T8.}
\end{figure}

\clearpage
\centerline{\ .}
\bigskip
\bigskip
\bigskip

\begin{figure}[h]
\figurenum{10}
\plotone{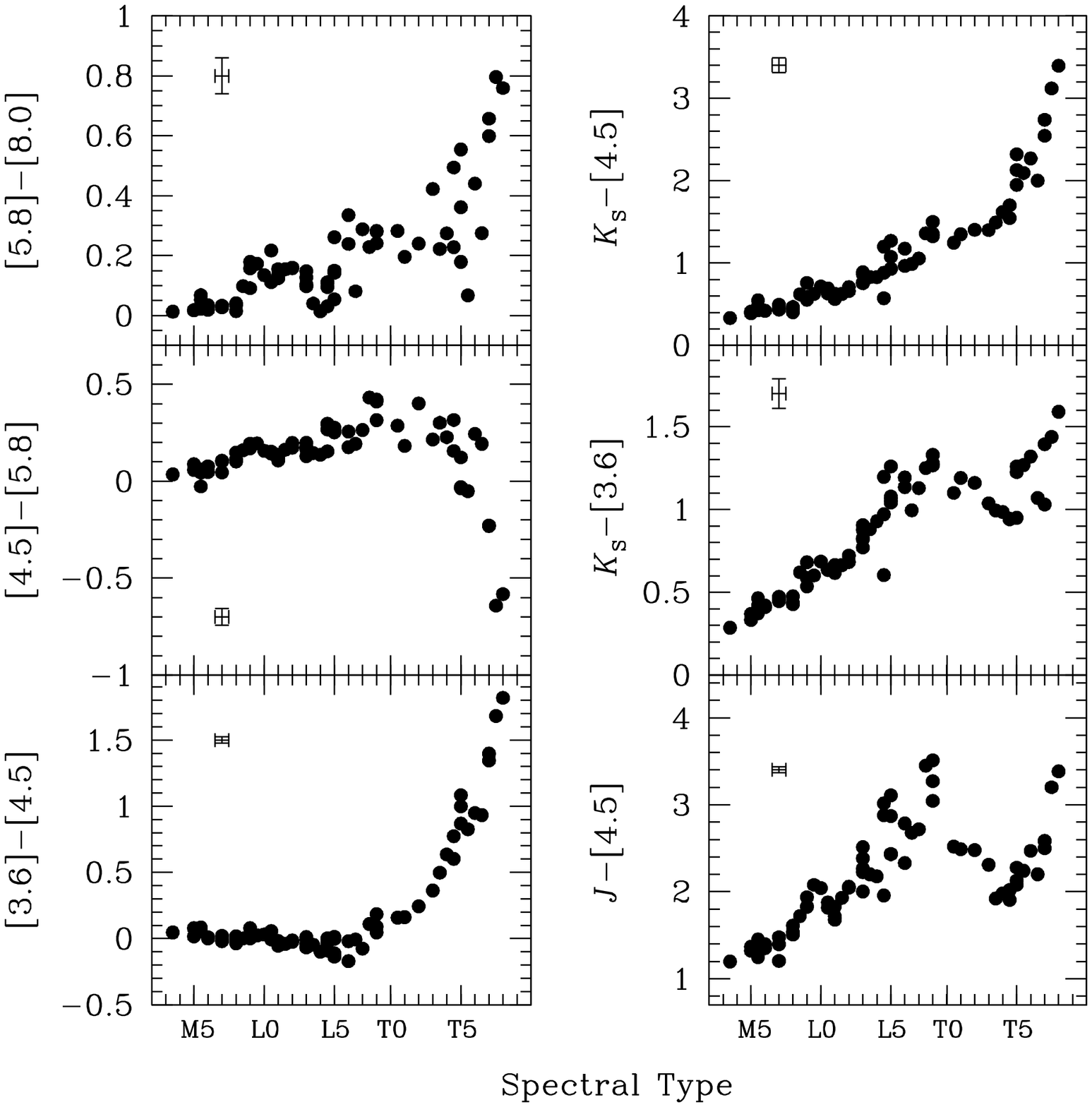}
\caption{A summary of the trending seen in various 
color indices versus spectral type in the same spirit as near-
and mid-IR plots shown in Leggett et al. (2002), Knapp et al. (2004), 
and Golimowski et al. (2004a).  The spectral types are the same as 
those used in Table 1, with optical types being used for the M and L 
dwarfs and infrared types for the T dwarfs.  All $J$- and $K$-band photometry 
are on
the 2MASS system, with all transformations made using the 
relations of Stephen \& Leggett (2005).  Representative error
bars for the median errors in the photometry that make up
each color index and an error of +/- one spectral subclass 
are shown in each plot.}
\end{figure}

\clearpage
\centerline{\ .}
\bigskip
\bigskip
\bigskip

\begin{figure}[h]
\figurenum{11}
%\plotone{./f11.eps}
\includegraphics[angle=270,scale=0.65]{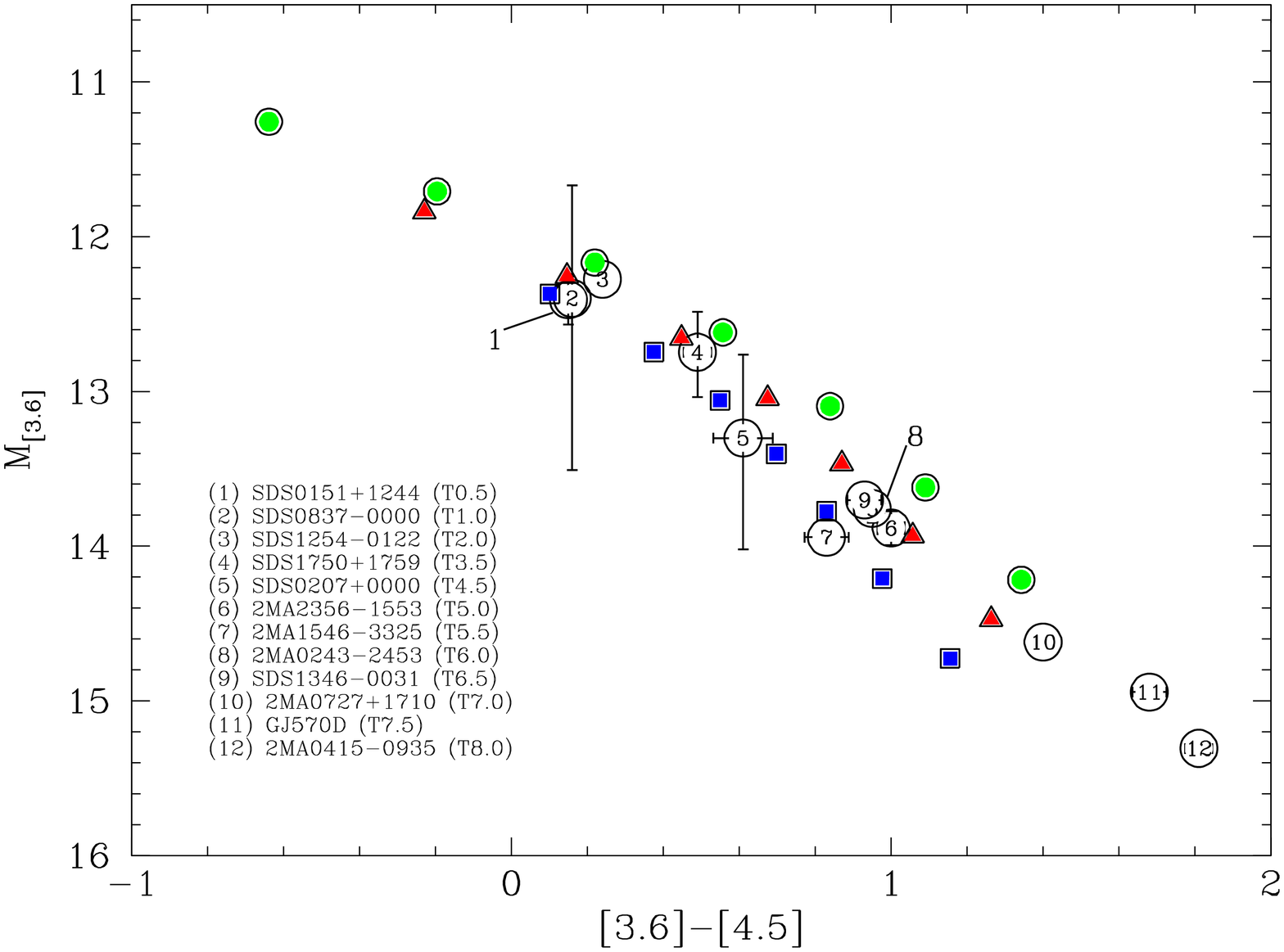}
\caption{A color-magnitude diagram of M$_{3.6}$ versus the 
[3.6]-[4.5] color for the T dwarfs in our program with trigonometric
parallaxes.  As in the previous figures in this paper, known binaries 
and spectrally peculiar objects have been excluded. 
Each T dwarf plotted is numbered 1 to 12 in order of increasing 
spectral subtype.  The accompanying error bars include the estimated
photometric errors and the published errors in the distance estimates
for each individual object (in some cases the error bars are smaller
than the plot symbol). 
The blue squares are models for \teff from
1300 to 700 K (left to right, in steps of 100 K) at g= 10$^{5.5}$ cm s$^{-2}$, 
the red triangles and green circles are are models for the same 
temperature range from at g= 10$^{5}$ cm s$^{-2}$ and g= 10$^{4.5}$ cm 
s$^{-2}$, respectively.  The corresponding masses
range from $\sim$15 M$_{Jupiter}$ to 70 M$_{Jupiter}$ (Burrows et al. 1997).}
\end{figure}

\clearpage
\centerline{\ .}
\bigskip
\bigskip
\bigskip

\begin{figure}[h]
\figurenum{12}
%\plotone{./f12.eps}
\includegraphics[angle=270,scale=0.65]{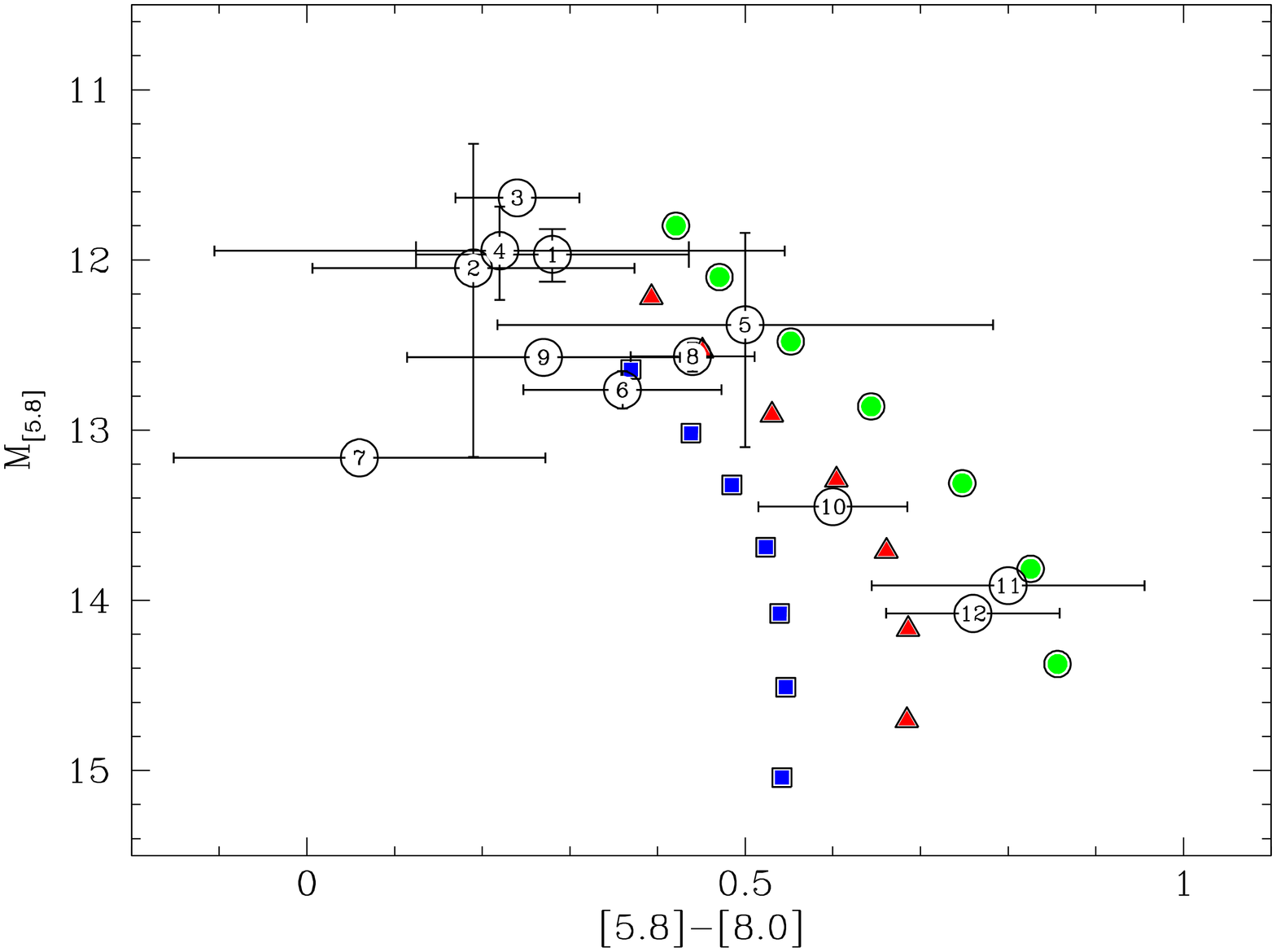}
\caption{A color-magnitude diagram of M$_{5.8}$ versus
the [5.8]-[8.0] color.  Otherwise, the format is the same as in Figure 11.  
The theoretical models are also the same and span \teff from 1300 to 700 K, 
in steps of 100 K, for gravities of 10$^{5.5}$ cm s$^{-2}$ ({\it blue 
squares}), 10$^{5}$ cm s$^{-2}$ ({\it red triangles}), and 10$^{4.5}$ cm 
s$^{-2}$ ({\it green circles}).}
\end{figure}

\clearpage
\centerline{\ .}
\bigskip
\bigskip
\bigskip

\begin{figure}[h]
\figurenum{13}
%\plotone{./f13.eps}
\includegraphics[angle=270,scale=0.65]{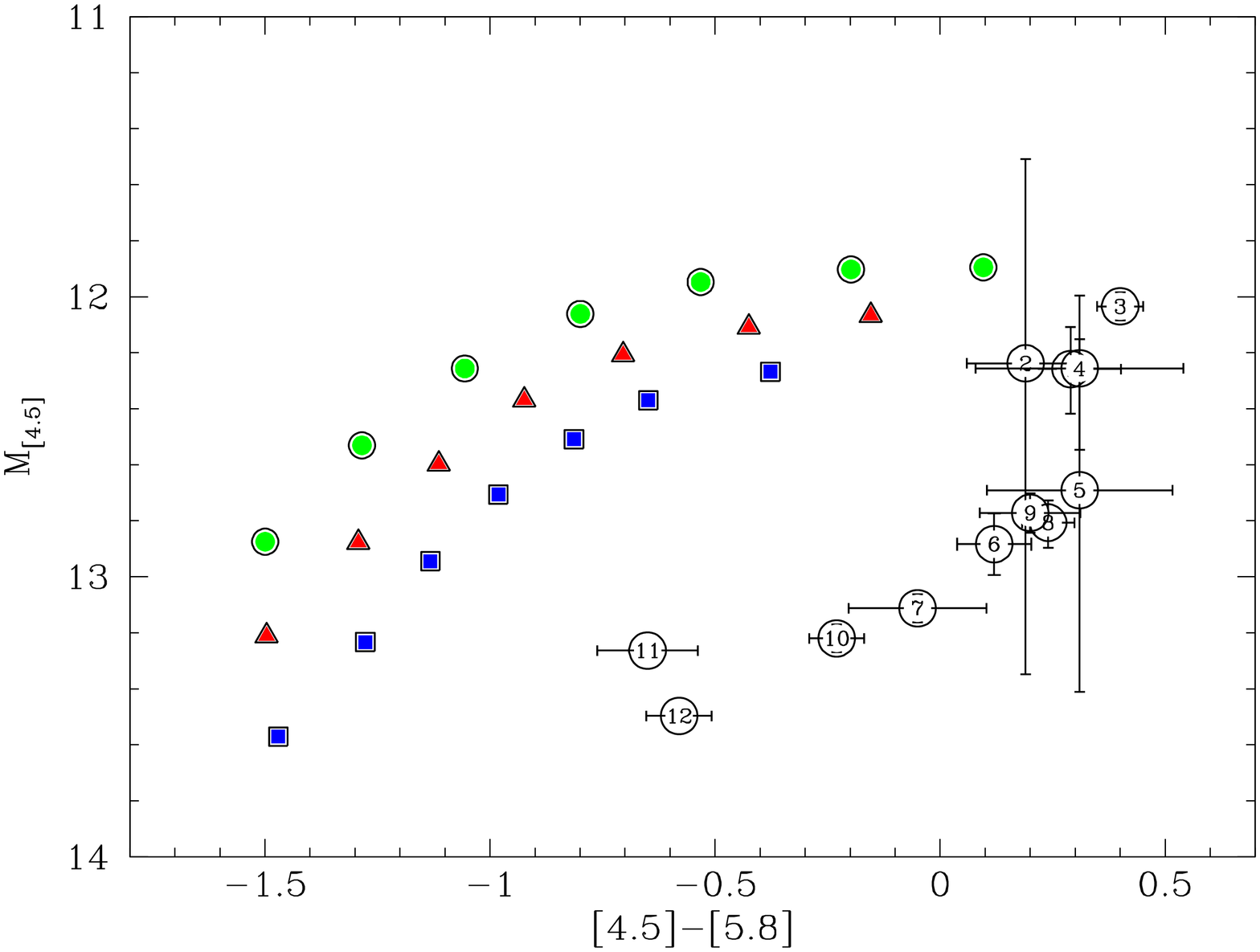}
\caption{A color-magnitude diagram of M$_{4.5}$ versus
the [4.5]-[5.8] color.  Otherwise, the format is the same as in Figure 11.  
For this color-magnitude diagram, the theoretical fits (the same as 
in Figure 11) are unacceptable.  
Because CO has a strong spectral feature at $\sim$4.67 \mic, one can interpret 
the discrepancy between theory and the IRAC data as an indication that 
equilibrium CH$_4$/CO chemistry underestimates the abundance of CO in 
T dwarf atmospheres (Golimowski et al. 2004a). The magnitude of the 
discrepancy translates into an overestimate in the $\sim$4--5 \mic flux 
by factors of 1.5 to 3.0.}
\end{figure}

\clearpage
\centerline{\ .}
\bigskip
\bigskip
\bigskip

\begin{figure}[h]
\figurenum{14}
%\plotone{./f14.eps}
\includegraphics[angle=270,scale=0.65]{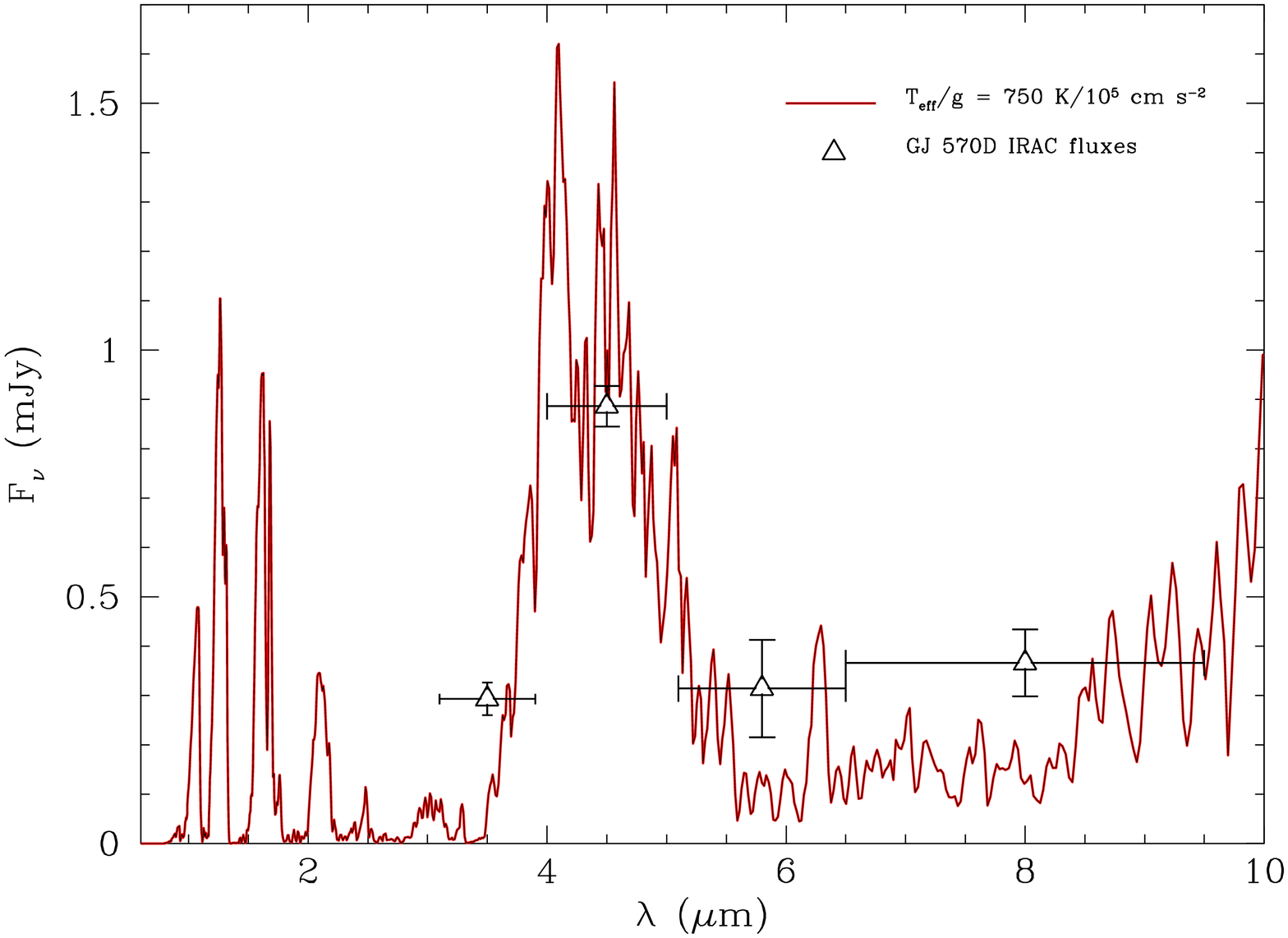}
\caption{A comparison between a theoretical T dwarf spectral model  
with \teff/g = 750 K/10$^{5}$ cm s$^{-2}$ and the IRAC fluxes measured
for the T7.5 dwarf GJ 570D.  A distance of 5.89 parsecs has been
assumed (Dahn et al. 2002).  The black triangles depict the fluxes (with 
their uncertainties) derived from the IRAC photometry; the horizontal lines
indicate the widths of the IRAC bands.  This model was generated for 
GJ 570D in 2002 to
fit its optical spectrum {\it shortward} of 1.0 \mic 
(Burrows et al. 2002).  No further attempt was made to
improve the fit.  A slight ($\sim$40\%) discrepancy in the 4.5-\mic bandpass, 
attributable to a CO abundance excess in the atmosphere, is manifest.  
Despite this, this plot and Figure 7 in Burrows et al. (2002), represent 
a good fit (that can nevertheless be improved) from 0.6 \mic to 
$\sim$8.0 \mic and indicate 
that the IRAC data were successfully anticipated.}
\end{figure}

%%
%%  Tables
%%
\clearpage
\LongTables
\begin{landscape}
%%  Table 1  Basic Data for MLT Dwarfs in the Sample
%%
%%
\begin{deluxetable}{lccccccrrrcc}
%\rotate
\tabletypesize{\scriptsize}
\tablewidth{617pt}
%\tablewidth{0pt}
\tablenum{1}
\tablecaption{The Sample of late-M, L, and T-type Dwarfs: Basic Data}
\tablecolumns{12}
\tablehead{
\colhead{} & \colhead{R.~A.} & \colhead{Dec.} & \colhead{Spectral} & \colhead{SpT} & \colhead{$\pi$~(error)} & \colhead{$\pi$} & \colhead{} & \colhead{} 
& \colhead{} & \colhead{$JHK$} & \colhead{Photometric}\\
\colhead{Name} & \colhead{(J2000.0)} & \colhead{(J2000.0)} & \colhead{Type\tablenotemark{a}} & \colhead{Ref} & \colhead{(arcseconds)} & \colhead{Ref} 
& \colhead{$J$~(error)} & \colhead{$H$~(error)} & \colhead{$K_S$~(error)} & \colhead{Ref} & \colhead{System}
}
\startdata
GJ1001A                           & 00 04 36.4 &$-$40 44 03 & M3.5       & 1 &     \nodata         & \nodata& 8.60  (0.01) & 8.04  (0.03) & 7.74  (0.04) & 1  & 2MA \\
GJ1093                            & 06 59 28.9 &  +19 20 53 & M5.0       & 2 & 0.12880 (0.00350)   & 1      & 9.16  (0.02) & 8.55  (0.02) & 8.23  (0.02) & 1  & 2MA \\
GJ1156                            & 12 18 59.5 &  +11 07 33 & M5.0       & 2 & 0.15290 (0.00300)   & 1      & 8.53  (0.03) & 7.88  (0.03) & 7.57  (0.03) & 1  & 2MA \\
GJ1002                            & 00 06 43.4 &$-$07 32 19 & M5.5       & 2 & 0.21300 (0.00360)   & 1      & 8.32  (0.02) & 7.79  (0.03) & 7.44  (0.02) & 1  & 2MA \\
LHS288                            & 10 44 21.3 &$-$61 12 35 & M5.5       & 3 & 0.22250 (0.01130)   & 1      & 8.49  (0.01) & 8.05  (0.04) & 7.73  (0.02) & 1  & 2MA \\
GJ412B                            & 11 05 30.3 &  +43 31 17 & M5.5       & 2 & 0.20694 (0.00119)   & 2      & 8.74  (0.03) & 8.18  (0.02) & 7.84  (0.03) & 1  & 2MA \\
GJ1111                            & 08 29 49.5 &  +26 46 32 & M6.5       & 4 & 0.27580 (0.00300)   & 1      & 8.24  (0.02) & 7.62  (0.02) & 7.26  (0.02) & 1  & 2MA \\
LHS292                            & 10 48 12.8 &$-$11 20 11 & M6.5       & 2 & 0.22030 (0.00360)   & 1      & 8.86  (0.02) & 8.26  (0.04) & 7.93  (0.03) & 1  & 2MA \\
SO0253+1652                       & 02 53 00.5 &  +16 52 58 & M7.0       & 5 & 0.26800 (0.03700)   & 3      & 8.39  (0.03) & 7.88  (0.04) & 7.59  (0.05) & 1  & 2MA \\ 
LHS3003                           & 14 56 38.4 &$-$28 09 48 & M7.0       & 6 & 0.15630 (0.00300)   & 1      & 9.97  (0.03) & 9.32  (0.02) & 8.93  (0.03) & 1  & 2MA \\
GJ644C                            & 16 55 35.3 &$-$08 23 40 & M7.0       & 4 & 0.15542 (0.00133)   & 4      & 9.78  (0.03) & 9.20  (0.02) & 8.82  (0.02) & 1  & 2MA \\   
LHS132                            & 01 02 51.2 &$-$37 37 45 & M8.0:      & 7 &     \nodata         & \nodata& 11.13 (0.02) & 10.48 (0.02) & 10.07 (0.02) & 1  & 2MA \\
LHS2021                           & 08 30 32.7 &  +09 47 14 & M8.0       & 5 &     \nodata         & \nodata& 11.89 (0.02) & 11.17 (0.02) & 10.76 (0.02) & 1  & 2MA \\
GJ752B                            & 19 16 57.7 &  +05 09 00 & M8.0       & 4 & 0.17010 (0.00080)   & 5      & 9.91  (0.03) & 9.23  (0.03) & 8.77  (0.02) & 1  & 2MA \\
2MA1835+3259                      & 18 35 37.9 &  +32 59 54 & M8.5       & 8 & 0.17650 (0.00050)   & 6      & 10.27 (0.02) & 9.62  (0.02) & 9.17  (0.02) & 1  & 2MA \\     
LP944-020                         & 03 39 35.2 &$-$35 25 44 & M9.0       & 9 & 0.20140 (0.00420)   & 7      & 10.73 (0.02) & 10.02 (0.02) & 9.55  (0.02) & 1  & 2MA \\
LHS2065                           & 08 53 36.1 &$-$03 29 30 & M9.0       & 4 & 0.11730 (0.00150)   & 1      & 11.21 (0.03) & 10.47 (0.03) & 9.94  (0.02) & 1  & 2MA \\
LHS2924                           & 14 28 42.0 &  +33 10 36 & M9.0       & 6 & 0.09080 (0.00130)   & 8      & 11.99 (0.02) & 11.23 (0.03) & 10.74 (0.02) & 1  & 2MA \\
DEN0021-4244                      & 00 21 05.8 &$-$42 44 49 & M9.5       &10 &     \nodata         & \nodata& 13.52 (0.03) & 12.81 (0.03) & 12.30 (0.03) & 1  & 2MA \\
BRI0021-0214                      & 00 24 24.6 &$-$01 58 20 & M9.5       & 1 & 0.08660 (0.00400)   & 8      & 11.99 (0.04) & 11.08 (0.02) & 10.54 (0.02) & 1  & 2MA \\
2MA1204+3212                      & 12 04 30.4 &  +32 13 00 & L0.0       &11 &     \nodata         & \nodata& 13.82 (0.04) & 13.09 (0.04) & 12.52 (0.03) & 1  & 2MA \\
2MA0320-0446                      & 03 20 28.4 &$-$04 46 36 & L0.5(IR)   &12 &     \nodata         & \nodata& 13.26 (0.02) & 12.54 (0.02) & 12.13 (0.03) & 1  & 2MA \\
2MA0451-3402                      & 04 51 00.9 &$-$34 02 15 & L0.5       &10 &     \nodata         & \nodata& 13.54 (0.02) & 12.83 (0.02) & 12.29 (0.03) & 1  & 2MA \\
2MA0746+2000AB\tablenotemark{b}   & 07 46 42.5 &  +20 00 32 & L0.5       &13 & 0.08190 (0.00030)   & 8      & 11.76 (0.02) & 11.01 (0.02) & 10.47 (0.02) & 1  & 2MA \\
2MA1300+1912                      & 13 00 42.6 &  +19 12 35 & L1.0       &14 &     \nodata         & \nodata& 12.72 (0.02) & 12.08 (0.02) & 11.62 (0.02) & 1  & 2MA \\
2MA1439+1929                      & 14 39 28.4 &  +19 29 15 & L1.0       &15 & 0.06960 (0.00050)   & 8      & 12.76 (0.02) & 12.04 (0.02) & 11.55 (0.02) & 1  & 2MA \\
2MA1555-0956                      & 15 55 15.7 &$-$09 56 06 & L1.0       &16 &     \nodata         & \nodata& 12.56 (0.02) & 11.98 (0.02) & 11.44 (0.03) & 1  & 2MA \\
2MA1645-1319                      & 16 45 22.1 &$-$13 19 52 & L1.5       &16 &     \nodata         & \nodata& 12.45 (0.03) & 11.69 (0.02) & 11.15 (0.03) & 1  & 2MA \\
2MA1017+1308                      & 10 17 07.5 &  +13 08 40 & L2.0:      &11 &     \nodata         & \nodata& 14.10 (0.02) & 13.28 (0.03) & 12.71 (0.02) & 1  & 2MA \\
2MA1155-3727                      & 11 55 39.5 &$-$37 27 35 & L2.0       &16 &     \nodata         & \nodata& 12.81 (0.02) & 12.04 (0.03) & 11.46 (0.02) & 1  & 2MA \\
Kelu-1\tablenotemark{b}           & 13 05 40.2 &$-$25 41 06 & L2.0       &15 & 0.05360 (0.00200)   & 8      & 13.41 (0.03) & 12.39 (0.03) & 11.75 (0.02) & 1  & 2MA \\
DEN1058-1548                      & 10 58 47.9 &$-$15 48 17 & L3.0       &15 & 0.05770 (0.00100)   & 8      & 14.16 (0.04) & 13.23 (0.03) & 12.53 (0.03) & 1  & 2MA \\
2MA1506+1321                      & 15 06 54.4 &  +13 21 06 & L3.0       &14 &     \nodata         & \nodata& 13.37 (0.02) & 12.38 (0.02) & 11.74 (0.02) & 1  & 2MA \\
2MA1721+3344                      & 17 21 03.9 &  +33 44 16 & L3.0       &11 &     \nodata         & \nodata& 13.63 (0.02) & 12.95 (0.03) & 12.49 (0.02) & 1  & 2MA \\
SDS2028+0052                      & 20 28 20.4 &  +00 52 27 & L3.0       &17 &     \nodata         & \nodata& 14.30 (0.04) & 13.38 (0.03) & 12.79 (0.03) & 1  & 2MA \\
2MA2104-1037                      & 21 04 14.9 &$-$10 37 37 & L3.0       &11 &     \nodata         & \nodata& 13.84 (0.03) & 12.98 (0.03) & 12.37 (0.02) & 1  & 2MA \\
2MA0036+1821                      & 00 36 15.9 &  +18 21 10 & L3.5       &13 & 0.11420 (0.00080)   & 8      & 12.32 (0.03) & 11.63 (0.03) & 11.03 (0.03) & 3  & MKO \\
DEN1539-0520                      & 15 39 41.9 &$-$05 20 43 & L4.0:      &18 &     \nodata         & \nodata& 13.92 (0.03) & 13.06 (0.03) & 12.58 (0.03) & 1  & 2MA \\
2MA0141+1804                      & 01 41 03.2 &  +18 04 50 & L4.5(IR)   &12 &     \nodata         & \nodata& 13.88 (0.03) & 13.03 (0.03) & 12.49 (0.03) & 1  & 2MA \\
2MA0652+4710                      & 06 52 30.7 &  +47 10 35 & L4.5       &11 &     \nodata         & \nodata& 13.51 (0.02) & 12.38 (0.02) & 11.69 (0.02) & 1  & 2MA \\
2MA2224-0158                      & 22 24 43.8 &$-$01 58 52 & L4.5       &19 & 0.08810 (0.00110)   & 8      & 13.89 (0.03) & 12.84 (0.03) & 11.98 (0.03) & 4  & MKO \\
GJ1001BC\tablenotemark{b}         & 00 04 34.9 &$-$40 44 06 & L5.0       &20 & 0.10470 (0.01140)   & 1      & 13.11 (0.02) & 12.06 (0.03) & 11.40 (0.03) & 1  & 2MA \\
SDS0539-0059                      & 05 39 52.0 &$-$00 59 02 & L5.0       &21 & 0.07612 (0.00217)   & 9      & 14.03 (0.03) & 13.10 (0.03) & 12.53 (0.02) & 1  & 2MA \\
2MA0835-0819                      & 08 35 42.6 &$-$08 19 24 & L5.0       &11 &     \nodata         & \nodata& 13.17 (0.02) & 11.94 (0.02) & 11.14 (0.02) & 1  & 2MA \\
2MA0908+5032                      & 09 08 38.0 &  +50 32 09 & L5.0       &11 &     \nodata         & \nodata& 14.40 (0.03) & 13.54 (0.03) & 12.89 (0.03) & 4  & MKO \\
2MA1507-1627                      & 15 07 47.6 &$-$16 27 38 & L5.0       &13 & 0.13640 (0.00060)   & 8      & 12.70 (0.03) & 11.80 (0.03) & 11.29 (0.03) & 3  & MKO \\
SDS1331-0116                      & 13 31 48.9 &$-$01 16 50 & L6.0       &17 &     \nodata         & \nodata& 15.32 (0.03) & 14.65 (0.03) & 14.07 (0.03) & 4  & MKO \\
2MA1515+4847                      & 15 15 00.8 &  +48 47 42 & L6.0(IR)   &12 &     \nodata         & \nodata& 14.11 (0.03) & 13.10 (0.03) & 12.50 (0.02) & 1  & 2MA \\
2MA0717+5705                      & 07 17 16.3 &  +57 05 43 & L6.5(IR)   &12 &     \nodata         & \nodata& 14.64 (0.03) & 13.59 (0.03) & 12.95 (0.03) & 1  & 2MA \\
2MA1526+2043                      & 15 26 14.1 &  +20 43 41 & L7.0       &19 &     \nodata         & \nodata& 15.59 (0.06) & 14.50 (0.04) & 13.92 (0.05) & 1  & 2MA \\
2MA1728+3948\tablenotemark{b}     & 17 28 11.5 &  +39 48 59 & L7.0       &19 & 0.04149 (0.00326)   & 9      & 15.99 (0.08) & 14.76 (0.07) & 13.91 (0.05) & 1  & 2MA \\
2MA0825+2115                      & 08 25 19.6 &  +21 15 52 & L7.5       &19 & 0.09381 (0.00100)   & 8      & 14.89 (0.03) & 13.81 (0.03) & 12.93 (0.03) & 3  & MKO \\
DEN0255-4700                      & 02 55 03.6 &$-$47 00 51 & L8.0       &18 &     \nodata         & \nodata& 13.25 (0.03) & 12.20 (0.02) & 11.56 (0.02) & 1  & 2MA \\
SDS0857+5708                      & 08 57 58.5 &  +57 08 51 & L8.0       &24 &     \nodata         & \nodata& 14.80 (0.03) & 13.80 (0.03) & 12.94 (0.03) & 3  & MKO \\
GJ337CD\tablenotemark{b}          & 09 12 14.5 &  +14 59 40 & L8.0       &23 & 0.04880 (0.00092)   & 8      & 15.51 (0.08) & 14.62 (0.08) & 14.04 (0.06) & 1  & 2MA \\   
2MA1632+1904                      & 16 32 29.1 &  +19 04 41 & L8.0       &15 & 0.06560 (0.00210)   & 8      & 15.77 (0.03) & 14.68 (0.03) & 13.97 (0.03) & 3  & MKO \\
2MA0532+8246                      & 05 32 53.5 &  +82 46 46 & sdL        &22 &     \nodata         & \nodata& 15.18 (0.06) & 14.90 (0.09) & 14.92 (0.15) & 1  & 2MA \\
SDS0423-0414\tablenotemark{b}     & 04 23 48.6 &$-$04 14 04 & T0.0       &25 & 0.06593 (0.00170)   & 9      & 14.30 (0.03) & 13.51 (0.03) & 12.96 (0.03) & 3  & MKO \\
SDS0151+1244                      & 01 51 41.7 &  +12 44 30 & T0.5       &25 & 0.04673 (0.00337)   & 9      & 16.25 (0.05) & 15.54 (0.05) & 15.18 (0.05) & 3  & MKO \\
SDS0837-0000                      & 08 37 17.2 &$-$00 00 18 & T1.0       &25 & 0.03370 (0.01345)   & 9      & 16.90 (0.05) & 16.21 (0.05) & 15.98 (0.05) & 5  & MKO \\
SDS1254-0122                      & 12 54 53.9 &$-$01 22 47 & T2.0       &25 & 0.08490 (0.00190)   & 8      & 14.66 (0.03) & 14.13 (0.03) & 13.84 (0.03) & 5  & MKO \\
SDS1021-0304\tablenotemark{b}     & 10 21 09.7 &$-$03 04 20 & T3.0       &25 & 0.03440 (0.00460)   & 7      & 15.88 (0.03) & 15.41 (0.03) & 15.26 (0.05) & 5  & MKO \\
SDS1750+1759                      & 17 50 33.0 &  +17 59 04 & T3.5       &25 & 0.03624 (0.00453)   & 9      & 16.14 (0.05) & 15.94 (0.05) & 16.02 (0.05) & 3  & MKO \\
2MA2254+3123                      & 22 54 18.8 &  +31 23 49 & T4.0       &25 &     \nodata         & \nodata& 15.26 (0.05) & 15.02 (0.08) & 14.90 (0.15) & 1  & 2MA \\
SDS0207+0000                      & 02 07 42.8 &  +00 00 56 & T4.5       &25 & 0.03485 (0.00987)   & 9      & 16.63 (0.05) & 16.66 (0.05) & 16.62 (0.05) & 3  & MKO \\
2MA0559-1404                      & 05 59 19.1 &$-$14 04 48 & T4.5       &25 & 0.09770 (0.00130)   & 8      & 13.57 (0.03) & 13.64 (0.03) & 13.73 (0.03) & 3  & MKO \\
SDS0926+5847\tablenotemark{b}     & 09 26 15.4 &  +58 47 21 & T4.5       &25 &     \nodata         & \nodata& 15.47 (0.03) & 15.42 (0.03) & 15.50 (0.03) & 3  & MKO \\
2MA0755+2212                      & 07 55 48.0 &  +22 12 18 & T5.0       &25 &     \nodata         & \nodata& 15.46 (0.03) & 15.70 (0.03) & 15.86 (0.03) & 4  & MKO \\
2MA2339+1352                      & 23 39 10.1 &  +13 52 30 & T5.0       &25 &     \nodata         & \nodata& 15.81 (0.03) & 16.00 (0.03) & 16.17 (0.03) & 4  & MKO \\
2MA2356-1553                      & 23 56 54.7 &$-$15 53 10 & T5.0       &25 & 0.06897 (0.00342)   & 9      & 15.48 (0.03) & 15.70 (0.03) & 15.73 (0.03) & 4  & MKO \\
2MA1534-2952\tablenotemark{b}     & 15 34 49.8 &$-$29 52 27 & T5.5       &25 & 0.07360 (0.00120)   & 7      & 14.60 (0.03) & 14.74 (0.03) & 14.91 (0.03) & 4  & MKO \\
2MA1546-3325                      & 15 46 27.1 &$-$33 25 11 & T5.5       &25 & 0.08800 (0.00190)   & 7      & 15.63 (0.05) & 15.45 (0.09) & 15.49 (0.18) & 1  & 2MA \\
SDS1110+0116                      & 11 10 10.0 &  +01 16 13 & T5.5       &25 &     \nodata         & \nodata& 16.12 (0.05) & 16.22 (0.05) & 16.05 (0.05) & 3  & MKO \\
2MA0243-2453                      & 02 43 13.7 &$-$24 53 29 & T6.0       &25 & 0.09362 (0.00363)   & 9      & 15.13 (0.03) & 15.39 (0.03) & 15.34 (0.03) & 4  & MKO \\
2MA1225-2739\tablenotemark{b}     & 12 25 54.3 &$-$27 39 47 & T6.0       &25 & 0.07510 (0.00250)   & 7      & 14.88 (0.03) & 15.17 (0.03) & 15.28 (0.03) & 3  & MKO \\
SDS1624+0029                      & 16 24 14.4 &  +00 29 16 & T6.0       &25 & 0.09090 (0.00120)   & 7      & 15.66 (0.05) & 15.83 (0.05) & 15.90 (0.11) & 6  & 2MA \\
2MA0937+2931                      & 09 37 34.7 &  +29 31 42 & T6p        &25 & 0.16284 (0.00388)   & 9      & 14.58 (0.04) & 14.67 (0.03) & 15.39 (0.06) & 4  & MKO \\     
2MA1047+2124                      & 10 47 53.9 &  +21 24 23 & T6.5       &25 & 0.09473 (0.00381)   & 9      & 15.77 (0.04) & 15.83 (0.03) & 16.20 (0.03) & 3  & MKO \\     
2MA1237+6526                      & 12 37 39.2 &  +65 26 15 & T6.5       &25 & 0.09607 (0.00478)   & 9      & 16.17 (0.05) & 16.21 (0.05) & 16.72 (0.06) & 6  & 2MA \\     
SDS1346-0031                      & 13 46 46.5 &$-$00 31 50 & T6.5       &25 & 0.06830 (0.00230)   & 7      & 15.49 (0.05) & 15.84 (0.05) & 15.73 (0.05) & 3  & MKO \\
2MA0727+1710                      & 07 27 18.2 &  +17 10 01 & T7.0       &25 & 0.11014 (0.00234)   & 9      & 15.19 (0.03) & 15.67 (0.03) & 15.69 (0.03) & 4  & MKO \\
2MA1553+1532\tablenotemark{b}     & 15 53 02.2 &  +15 32 36 & T7.0       &25 &     \nodata         & \nodata& 15.34 (0.03) & 15.76 (0.03) & 15.95 (0.03) & 4  & MKO \\
2MA1217-0311                      & 12 17 11.1 &$-$03 11 13 & T7.5       &25 & 0.09080 (0.00220)   & 7      & 15.56 (0.03) & 15.98 (0.03) & 15.92 (0.03) & 3  & MKO \\     
GJ570D                            & 14 57 15.0 &$-$21 21 48 & T7.5       &25 & 0.16930 (0.00170)   & 10     & 15.32 (0.05) & 15.27 (0.09) & 15.24 (0.16) & 1  & 2MA \\
2MA0415-0935                      & 04 15 19.5 &$-$09 35 06 & T8.0       &25 & 0.17434 (0.00276)   & 9      & 15.32 (0.03) & 15.70 (0.04) & 15.83 (0.03) & 4  & MKO \\
$\epsilon$ Ind BC\tablenotemark{b}& 22 04 10.5 &$-$56 46 58 & T1.0+T6.0  &25 & 0.27580 (0.00069)   & 10     & 11.91 (0.02) & 11.31 (0.02) & 11.21 (0.02) & 1  & 2MA \\  
\enddata
\tablerefs{SPECTRAL TYPE REFERENCES $-$ 
(1) Golimowski et al. 2004, 
(2) Henry, Kirkpatrick, \& Simons 1994
(3) Bessell 1991, 
(4) Kirkpatrick, Henry, \& McCarthy 1991, 
(5) Henry et al. 2004,
(6) Kirkpatrick, Henry, \& Simons 1995, 
(7) Scholz et al. 2000,
(8) Reid 2003,
(9) Kirkpatrick, Henry, \& Irwin 1997,
(10) Basri et al. 2000,
(11) Cruz et al. 2003,
(12) Wilson et al. 2003,
(13) Reid et al. 2000,
(14) Gizis et al. 2000,
(15) Kirkpatrick et al. 1999,
(16) Gizis 2002,
(17) Hawley et al. 2002,
(18) Kirkpatrick et al. 2006,
(19) Kirkpatrick et al. 2000,
(20) Kirkpatrick et al. 2001,
(21) Fan et al. 2000,
(22) Burgasser et al. 2003a,
(23) Wilson et al. 2001,
(24) Geballe et al. 2002,
(25) Burgasser et al. 2005.
~~PARALLAX REFERENCES $-$
(1) van Altena, Lee, \& Hoffleit 1995,
(2) Gould \& Chaname 2004,
(3) Henry et al. 2004,
(4) Costa et al. 2005,
(5) Reid \& Gizis 2005,
(6) Reid et al. 2003,
(7) Tinney, Burgasser, \& Kirkpatrick 2003,
(8) Dahn et al. 2002,
(9) Vrba et al. 2004,
(10) Perryman et al. 1997.
~~PHOTOMETRY REFERENCES $-$
(1) 2MASS,
(2) Leggett 1992,
(3) Leggett et al. 2002,
(4) Knapp et al. 2004,
(5) Leggett et al. 2000,
(6) this paper.
}
\tablenotetext{a}{Unless otherwise noted, optical spectral types are used for M and L dwarfs 
and infrared spectral types are used for the T dwarfs.}
\tablenotetext{b}{Known binary.}
\end{deluxetable}

\clearpage
\end{landscape}

%\clearpage
%\LongTables
%\begin{landscape}
%\input{tab2.tex}
%\clearpage
%\end{landscape}

\clearpage
\centerline{\ }
\bigskip
\bigskip
%%%%%%%%%%%%%%%%%%%%%%%%%%%%%%%%%%%%%%%%%%%%%%%%%%%%%%%%%%%%%%%%%%%%%%%%%%%%%%%
%
%
% Table 2  Calibration Parameters
%
\begin{deluxetable}{lcccc}
%\rotate
\tabletypesize{\footnotesize}
\tablewidth{0pt}
\tablenum{2}
\tablecaption{IRAC Photometry Calibration}
\tablecolumns{5}
\tablehead{
\colhead{} & \multicolumn{4}{c}{IRAC Channel} \\
\cline{2-5}
\colhead{Item} & \colhead{1 (3.6$\mu$m)} & \colhead{2 (4.5$\mu$m)} & \colhead{3 (5.8$\mu$m)} & \colhead{4 (8.0$\mu$m)} 
}
\startdata
Nominal $\lambda$~~[$\mu$m]                 & 3.550 & 4.493 & 5.731 & 7.872 \\
BCD calibration~~[(MJy/sr)/(DN/s)]            & 0.1088 & 0.1388 & 0.5952 & 0.2021 \\
Point source calibration~~[$\mu$Jy/(DN/s)]    & 3.813 & 4.800 & 20.891 & 7.070 \\
Aperture correction(error) for 4-pixel radius~~[mag]~~~~     & -0.084 (0.009)&  -0.089 (0.006)&  -0.072 (0.015)&  -0.077 (0.034) \\
$F_{\nu}$(Vega)~~[Jy]                         & 280.9 & 179.7 & 115.0 & 64.13 \\ 
\enddata
\end{deluxetable}

\clearpage
\LongTables
\begin{landscape}
%%%%%%%%%%%%%%%%%%%%%%%%%%%%%%%%%%%%%%%%%%%%%%%%%%%%%%%%%%%%%%%%%%%%%%%%%%%%%
%
%
%  Table 3  IRAC Phot and Colors for MLT Dwarfs
%
\begin{deluxetable}{lcrcrcrcrccccc}
%\rotate
\tabletypesize{\scriptsize}
\tablewidth{601pt}
%\tablewidth{0pt}
\tablenum{3}
\tablecaption{IRAC Photometry and Colors of late-M, L, and T Dwarfs}
\tablecolumns{14}
\tablehead{
\colhead{} & \colhead{Spectral} & \colhead{} & \colhead{} & \colhead{} & \colhead{} & \colhead{} 
& \colhead{} & \colhead{} & \colhead{} & \colhead{} & \colhead{} & \colhead{} & \colhead{} \\
\colhead{Name} & \colhead{Type} & \colhead{[3.6]~(error)} & \colhead{n} & \colhead{[4.5]~(error)} & \colhead{n} 
& \colhead{[5.8]~(error)} & \colhead{n} & \colhead{[8.0]~(error)} & \colhead{n} & \colhead{[3.6]-[4.5]} & \colhead{[4.5]-[5.8]}   
& \colhead{[5.8]-[8.0]} & \colhead{Notes}   
}
\startdata
GJ1001A                           & M3.5       & 7.45    (0.03)& 5  & 7.40    (0.03)& 5   & 7.37    (0.01)& 5   & 7.36    (0.01)& 5  &\phs0.05    &\phs0.04   & 0.01  & 1      \\
GJ1093                            & M5.0       & 7.86    (0.03)& 4  & 7.84    (0.02)& 5   & 7.76    (0.01)& 5   & 7.74    (0.01)& 5  &\phs0.02    &\phs0.09   & 0.02  & 1          \\
GJ1156                            & M5.0       & 7.24    (0.03)& 5  & 7.16    (0.02)& 5   & 7.10    (0.01)& 5   & 7.08    (0.01)& 5  &\phs0.08    &\phs0.06   & 0.02  & 1          \\
GJ1002                            & M5.5       & 7.07    (0.01)& 5  & 7.01    (0.01)& 5   & 6.97    (0.02)& 5   & 6.95    (0.01)& 5  &\phs0.05    &\phs0.04   & 0.02  & 2  \\
LHS288                            & M5.5       & 7.31    (0.03)& 5  & 7.25    (0.04)& 5   & 7.27    (0.01)& 5   & 7.20    (0.01)& 5  &\phs0.06    &$-$0.03    & 0.07  & 2  \\
GJ412B                            & M5.5       & 7.38    (0.01)& 5  & 7.29    (0.05)& 5   & 7.23    (0.02)& 5   & 7.18    (0.00)& 5  &\phs0.08    &\phs0.06   & 0.05  & 2          \\
GJ1111                            & M6.5       & 6.84    (0.02)& 5  & 6.84    (0.04)& 5   & 6.76    (0.05)& 5   & 6.74    (0.01)& 5  &\phs0.00    &\phs0.08   & 0.02  & 3          \\
LHS292                            & M6.5       & 7.52    (0.02)& 5  & 7.51    (0.02)& 5   & 7.46    (0.02)& 5   & 7.42    (0.01)& 5  &\phs0.01    &\phs0.05   & 0.03  & 2          \\
SO0253+1652                       & M7.0       & 7.12    (0.01)& 3  & 7.10    (0.02)& 5   & 7.05    (0.01)& 5   & 7.02    (0.01)& 5  &\phs0.02    &\phs0.04   & 0.03  & 2          \\
LHS3003                           & M7.0       & 8.47    (0.02)& 5  & 8.49    (0.01)& 5   & 8.39    (0.02)& 5   & 8.36    (0.01)& 5  &$-$0.02     &\phs0.10   & 0.03  & 4,5        \\
GJ644C                            & M7.0       & 8.37    (0.02)& 5  & 8.38    (0.01)& 5   & 8.28    (0.02)& 5   & 8.24    (0.02)& 5  &$-$0.01     &\phs0.11   & 0.03  & 4,5        \\
LHS132                            & M8.0:      & 9.64    (0.02)& 5  & 9.62    (0.02)& 5   & 9.52    (0.02)& 5   & 9.48    (0.01)& 4  &\phs0.02    &\phs0.10   & 0.04  & 4,5        \\
LHS2021                           & M8.0       & 10.32   (0.02)& 5  & 10.35   (0.01)& 5   & 10.24   (0.01)& 5   & 10.20   (0.01)& 5  &$-$0.04     &\phs0.11   & 0.03  & 4          \\
GJ752B                            & M8.0       & 8.29    (0.02)& 5  & 8.30    (0.03)& 5   & 8.15    (0.01)& 5   & 8.14    (0.00)& 5  &$-$0.01     &\phs0.15   & 0.02  & 4,5        \\
2MA1835+3259                      & M8.5       & 8.55    (0.02)& 5  & 8.55    (0.01)& 5   & 8.39    (0.01)& 5   & 8.29    (0.01)& 5  &\phs0.00    &\phs0.16   & 0.10  & 4,5        \\
LP944-020                         & M9.0       & 8.87    (0.03)& 5  & 8.79    (0.01)& 5   & 8.59    (0.01)& 5   & 8.42    (0.01)& 5  &\phs0.08    &\phs0.19   & 0.18  & 4,5,6 \\
LHS2065                           & M9.0       & 9.41    (0.02)& 5  & 9.39    (0.03)& 5   & 9.22    (0.01)& 5   & 9.13    (0.01)& 5  &\phs0.02    &\phs0.17   & 0.09  & 4,5        \\
LHS2924                           & M9.0       & 10.16   (0.02)& 5  & 10.16   (0.01)& 5   & 9.97    (0.01)& 5   & 9.81    (0.01)& 5  &\phs0.00    &\phs0.19   & 0.16  & 4          \\
DEN0021-4244                      & M9.5       & 11.62   (0.01)& 5  & 11.59   (0.01)& 5   & 11.43   (0.04)& 5   & 11.30   (0.04)& 5  &\phs0.03    &\phs0.16   & 0.14  & \nodata\\
BRI0021-0214                      & M9.5       & 9.94    (0.03)& 4  & 9.91    (0.03)& 4   & 9.72    (0.01)& 5   & 9.55    (0.01)& 4  &\phs0.02    &\phs0.20   & 0.17  & 4,5        \\
2MA1204+3212                      & L0.0       & 11.93   (0.01)& 5  & 11.95   (0.01)& 5   & 11.82   (0.01)& 5   & 11.65   (0.03)& 5  &$-$0.01     &\phs0.13   & 0.17  & \nodata\\
2MA0320-0446                      & L0.5(IR)   & 11.50   (0.03)& 5  & 11.44   (0.01)& 5   & 11.29   (0.02)& 5   & 11.18   (0.01)& 4  &\phs0.06    &\phs0.15   & 0.11  & \nodata\\
2MA0451-3402                      & L0.5       & 11.66   (0.04)& 5  & 11.66   (0.03)& 5   & 11.52   (0.02)& 5   & 11.30   (0.04)& 5  &$-$0.01     &\phs0.14   & 0.22  & 7          \\
2MA0746+2000AB\tablenotemark{a}   & L0.5       & 9.86    (0.02)& 5  & 9.90    (0.04)& 4   & 9.72    (0.01)& 4   & 9.57    (0.01)& 5  &$-$0.03     &\phs0.18   & 0.15  & 4,5        \\
2MA1300+1912                      & L1.0       & 10.96   (0.02)& 5  & 11.00   (0.03)& 5   & 10.86   (0.01)& 5   & 10.73   (0.03)& 5  &$-$0.04     &\phs0.14   & 0.14  & \nodata\\
2MA1439+1929                      & L1.0       & 10.91   (0.02)& 5  & 10.93   (0.03)& 5   & 10.82   (0.03)& 5   & 10.67   (0.02)& 5  &$-$0.02     &\phs0.11   & 0.15  & 4,5        \\
2MA1555-0956                      & L1.0       & 10.83   (0.01)& 5  & 10.88   (0.01)& 5   & 10.76   (0.02)& 5   & 10.63   (0.01)& 5  &$-$0.05     &\phs0.12   & 0.12  & \nodata\\
2MA1645-1319                      & L1.5       & 10.48   (0.04)& 5  & 10.52   (0.02)& 5   & 10.36   (0.01)& 5   & 10.20   (0.01)& 5  &$-$0.04     &\phs0.16   & 0.15  & 4          \\
2MA1017+1308                      & L2.0:      & 12.03   (0.01)& 5  & 12.05   (0.03)& 5   & 11.85   (0.04)& 4   & 11.70   (0.03)& 5  &$-$0.02     &\phs0.20   & 0.16  & \nodata\\
2MA1155-3727                      & L2.0       & 10.74   (0.02)& 5  & 10.75   (0.02)& 5   & 10.58   (0.01)& 5   & 10.42   (0.02)& 5  &$-$0.01     &\phs0.17   & 0.16  & 4          \\
Kelu-1\tablenotemark{a}           & L2.0       & 10.92   (0.05)& 5  & 10.90   (0.04)& 5   & 10.73   (0.01)& 5   & 10.61   (0.02)& 5  &\phs0.02    &\phs0.17   & 0.12  & 4,5        \\
DEN1058-1548                      & L3.0       & 11.76   (0.02)& 5  & 11.77   (0.02)& 5   & 11.60   (0.02)& 5   & 11.50   (0.02)& 5  &$-$0.01     &\phs0.17   & 0.10  & \nodata\\
2MA1506+1321                      & L3.0       & 10.86   (0.02)& 5  & 10.85   (0.06)& 5   & 10.69   (0.02)& 5   & 10.58   (0.01)& 5  &\phs0.01    &\phs0.17   & 0.10  & 4,5        \\
2MA1721+3344                      & L3.0       & 11.58   (0.02)& 5  & 11.62   (0.02)& 5   & 11.49   (0.04)& 5   & 11.40   (0.02)& 5  &$-$0.04     &\phs0.13   & 0.10  & 8          \\
SDS2028+0052                      & L3.0       & 11.97   (0.02)& 5  & 12.03   (0.02)& 5   & 11.83   (0.03)& 5   & 11.71   (0.03)& 5  &$-$0.06     &\phs0.20   & 0.13  & \nodata\\
2MA2104-1037                      & L3.0       & 11.55   (0.03)& 5  & 11.62   (0.01)& 5   & 11.44   (0.03)& 5   & 11.29   (0.04)& 4  &$-$0.07     &\phs0.18   & 0.15  & \nodata\\
2MA0036+1821                      & L3.5       & 10.19   (0.03)& 5  & 10.24   (0.01)& 5   & 10.10   (0.02)& 5   & 10.06   (0.01)& 5  &$-$0.05     &\phs0.15   & 0.04  & 4          \\
DEN1539-0520                      & L4.0:      & 11.65   (0.02)& 5  & 11.75   (0.04)& 5   & 11.61   (0.05)& 5   & 11.60   (0.05)& 5  &$-$0.10     &\phs0.14   & 0.01  & \nodata\\
2MA0141+1804                      & L4.5(IR)   & 11.89   (0.04)& 4  & 11.92   (0.01)& 5   & 11.76   (0.04)& 5   & 11.67   (0.03)& 5  &$-$0.03     &\phs0.15   & 0.09  & 9          \\
2MA0652+4710                      & L4.5       & 10.50   (0.01)& 5  & 10.50   (0.01)& 5   & 10.23   (0.01)& 5   & 10.12   (0.02)& 5  &\phs0.00    &\phs0.27   & 0.11  & 4          \\
2MA2224-0158                      & L4.5       & 11.05   (0.02)& 5  & 11.14   (0.02)& 5   & 10.85   (0.01)& 5   & 10.81   (0.02)& 5  &$-$0.09     &\phs0.30   & 0.03  & \nodata\\
GJ1001BC\tablenotemark{a}         & L5.0       & 10.36   (0.01)& 5  & 10.47   (0.01)& 5   & 10.14   (0.03)& 5   & 10.13   (0.02)& 5  &$-$0.11     &\phs0.33   & 0.01  & 10         \\
SDS0539-0059                      & L5.0       & 11.49   (0.02)& 5  & 11.60   (0.02)& 5   & 11.35   (0.03)& 4   & 11.20   (0.04)& 4  &$-$0.11     &\phs0.25   & 0.14  & 11         \\
2MA0835-0819                      & L5.0       & 10.06   (0.03)& 5  & 10.06   (0.02)& 5   & 9.79    (0.01)& 5   & 9.73    (0.00)& 5  &\phs0.00    &\phs0.28   & 0.05  & 4          \\
2MA0908+5032                      & L5.0       & 11.67   (0.02)& 5  & 11.66   (0.01)& 5   & 11.39   (0.01)& 4   & 11.13   (0.03)& 5  &\phs0.01    &\phs0.27   & 0.26  & \nodata\\
2MA1507-1627                      & L5.0       & 10.27   (0.03)& 5  & 10.40   (0.02)& 5   & 10.14   (0.02)& 5   & 9.99    (0.01)& 5  &$-$0.14     &\phs0.26   & 0.15  & 4,5        \\
SDS1331-0116                      & L6.0       & 12.96   (0.02)& 4  & 13.13   (0.02)& 5   & 12.95   (0.08)& 4   & 12.62   (0.06)& 4  &$-$0.17     &\phs0.18   & 0.33  & \nodata\\
2MA1515+4847                      & L6.0(IR)   & 11.31   (0.02)& 5  & 11.33   (0.02)& 5   & 11.07   (0.02)& 5   & 10.83   (0.02)& 5  &$-$0.02     &\phs0.26   & 0.24  & \nodata\\
2MA0717+5705                      & L6.5(IR)   & 11.95   (0.01)& 4  & 11.96   (0.01)& 5   & 11.76   (0.01)& 5   & 11.68   (0.01)& 4  &$-$0.01     &\phs0.19   & 0.08  & \nodata\\
2MA1526+2043                      & L7.0       & 12.79   (0.02)& 5  & 12.87   (0.03)& 5   & 12.60   (0.11)& 5   & 12.32   (0.04)& 5  &$-$0.07     &\phs0.26   & 0.29  & \nodata\\
2MA1728+3948\tablenotemark{a}     & L7.0       & 12.72   (0.02)& 5  & 12.66   (0.01)& 5   & 12.29   (0.04)& 4   & 12.13   (0.03)& 5  &\phs0.06    &\phs0.37   & 0.15  & \nodata\\
2MA0825+2115                      & L7.5       & 11.70   (0.03)& 5  & 11.59   (0.01)& 5   & 11.16   (0.01)& 4   & 10.93   (0.02)& 5  &\phs0.11    &\phs0.43   & 0.23  & \nodata\\
DEN0255-4700                      & L8.0       & 10.29   (0.02)& 5  & 10.20   (0.02)& 5   & 9.89    (0.01)& 5   & 9.61    (0.01)& 5  &\phs0.09    &\phs0.32   & 0.28  & 4,5        \\
SDS0857+5708                      & L8.0       & 11.62   (0.00)& 4  & 11.44   (0.02)& 5   & 11.02   (0.01)& 5   & 10.74   (0.02)& 5  &\phs0.19    &\phs0.42   & 0.28  & \nodata\\
GJ337CD\tablenotemark{a}          & L8.0       & 12.50   (0.02)& 4  & 12.33   (0.02)& 4   & 11.96   (0.08)& 4   & 11.95   (0.05)& 4  &\phs0.18    &\phs0.36   & 0.02  & \nodata\\
2MA1632+1904                      & L8.0       & 12.70   (0.03)& 5  & 12.65   (0.02)& 5   & 12.24   (0.04)& 5   & 12.00   (0.04)& 4  &\phs0.05    &\phs0.41   & 0.24  & \nodata\\
2MA0532+8246                      & sdL        & 13.37   (0.03)& 5  & 13.22   (0.02)& 5   & 13.23   (0.10)& 4   & 13.03   (0.10)& 4  &\phs0.15    &$-$0.02    & 0.20  & \nodata\\
SDS0423-0414\tablenotemark{a}     & T0.0       & 11.73   (0.02)& 4  & 11.58   (0.02)& 5   & 11.30   (0.01)& 5   & 11.01   (0.03)& 5  &\phs0.14    &\phs0.29   & 0.28  & \nodata\\
SDS0151+1244                      & T0.5       & 14.06   (0.02)& 5  & 13.91   (0.02)& 5   & 13.62   (0.11)& 5   & 13.34   (0.18)& 4  &\phs0.16    &\phs0.29   & 0.28  & \nodata   \\
SDS0837-0000                      & T1.0       & 14.76   (0.03)& 5  & 14.60   (0.01)& 5   & 14.41   (0.13)& 5   & 14.22   (0.14)& 3  &\phs0.16    &\phs0.18   & 0.20  & \nodata   \\
SDS1254-0122                      & T2.0       & 12.63   (0.01)& 5  & 12.39   (0.01)& 4   & 11.99   (0.05)& 5   & 11.75   (0.04)& 5  &\phs0.24    &\phs0.40   & 0.24  & \nodata\\
SDS1021-0304\tablenotemark{a}     & T3.0       & 14.16   (0.02)& 5  & 13.80   (0.02)& 5   & 13.58   (0.12)& 5   & 13.16   (0.11)& 5  &\phs0.36    &\phs0.22   & 0.42  & \nodata   \\
SDS1750+1759                      & T3.5       & 14.95   (0.03)& 5  & 14.46   (0.02)& 5   & 14.15   (0.23)& 5   & 13.93   (0.23)& 5  &\phs0.50    &\phs0.30   & 0.22  & \nodata   \\
2MA2254+3123                      & T4.0       & 13.92   (0.03)& 5  & 13.28   (0.01)& 5   & 13.05   (0.10)& 5   & 12.78   (0.10)& 5  &\phs0.64    &\phs0.23   & 0.27  & \nodata   \\
SDS0207+0000                      & T4.5       & 15.59   (0.06)& 5  & 14.98   (0.05)& 5   & 14.67   (0.20)& 4   & 14.17   (0.19)& 4  &\phs0.60    &\phs0.32   & 0.49  & \nodata   \\
2MA0559-1404                      & T4.5       & 12.67   (0.03)& 5  & 11.93   (0.02)& 5   & 11.73   (0.02)& 5   & 11.42   (0.02)& 5  &\phs0.75    &\phs0.20   & 0.31  & \nodata\\
SDS0926+5847\tablenotemark{a}     & T4.5       & 14.48   (0.03)& 5  & 13.71   (0.02)& 5   & 13.55   (0.11)& 5   & 13.32   (0.06)& 4  &\phs0.77    &\phs0.16   & 0.23  & \nodata   \\
2MA0755+2212                      & T5.0       & 14.54   (0.03)& 5  & 13.45   (0.03)& 5   & 13.48   (0.10)& 5   & 12.93   (0.22)& 4  &\phs1.08    &$-$0.03    & 0.55  & \nodata   \\
2MA2339+1352                      & T5.0       & 14.82   (0.04)& 5  & 13.95   (0.04)& 5   & 13.98   (0.03)& 4   & 13.80   (0.20)& 4  &\phs0.87    &$-$0.04    & 0.18  & \nodata   \\
2MA2356-1553                      & T5.0       & 14.69   (0.03)& 5  & 13.69   (0.02)& 5   & 13.57   (0.08)& 5   & 13.21   (0.17)& 5  &\phs1.00    &\phs0.12   & 0.36  & \nodata   \\
2MA1534-2952\tablenotemark{a}     & T5.5       & 13.63   (0.04)& 5  & 12.71   (0.02)& 5   & 12.73   (0.05)& 5   & 12.36   (0.08)& 5  &\phs0.92    &$-$0.02    & 0.37  & 8          \\
2MA1546-3325                      & T5.5       & 14.22   (0.05)& 5  & 13.39   (0.03)& 5   & 13.44   (0.15)& 4   & 13.38   (0.10)& 4  &\phs0.83    &$-$0.05    & 0.07  & \nodata   \\
SDS1110+0116                      & T5.5       & 14.71   (0.03)& 4  & 13.88   (0.02)& 4   & 13.43   (0.07)& 5   & 13.21   (0.16)& 5  &\phs0.83    &\phs0.45   & 0.22  & \nodata   \\
2MA0243-2453                      & T6.0       & 13.90   (0.01)& 5  & 12.95   (0.03)& 5   & 12.71   (0.05)& 4   & 12.27   (0.05)& 4  &\phs0.95    &\phs0.24   & 0.44  & \nodata\\
2MA1225-2739\tablenotemark{a}     & T6.0       & 13.84   (0.02)& 5  & 12.75   (0.01)& 5   & 12.84   (0.10)& 5   & 12.24   (0.02)& 5  &\phs1.09    &$-$0.09    & 0.60  & \nodata\\
SDS1624+0029                      & T6.0       & 14.30   (0.03)& 5  & 13.08   (0.02)& 5   & 13.25   (0.08)& 5   & 12.84   (0.09)& 5  &\phs1.22    &$-$0.17    & 0.41  & \nodata\\
2MA0937+2931                      & T6p        & 13.10   (0.03)& 5  & 11.64   (0.04)& 5   & 12.32   (0.02)& 4   & 11.73   (0.04)& 5  &\phs1.47    &$-$0.68    & 0.58  & \nodata\\
2MA1047+2124                      & T6.5       & 14.39   (0.06)& 5  & 12.95   (0.04)& 5   & 13.52   (0.07)& 5   & 12.91   (0.10)& 5  &\phs1.44    &$-$0.57    & 0.61  & \nodata   \\
2MA1237+6526                      & T6.5       & 14.39   (0.03)& 4  & 12.93   (0.03)& 5   & 13.42   (0.06)& 5   & 12.78   (0.11)& 5  &\phs1.45    &$-$0.49    & 0.65  & \nodata\\
SDS1346-0031                      & T6.5       & 14.53   (0.04)& 5  & 13.60   (0.02)& 5   & 13.40   (0.11)& 5   & 13.13   (0.17)& 5  &\phs0.93    &\phs0.19   & 0.28  & \nodata   \\
2MA0727+1710                      & T7.0       & 14.41   (0.02)& 5  & 13.01   (0.01)& 5   & 13.24   (0.06)& 5   & 12.64   (0.11)& 5  &\phs1.40    &$-$0.23    & 0.60  & \nodata   \\
2MA1553+1532\tablenotemark{a}     & T7.0       & 14.42   (0.01)& 5  & 13.08   (0.02)& 3   & 13.30   (0.10)& 5   & 12.65   (0.10)& 4  &\phs1.35    &$-$0.23    & 0.66  & \nodata\\
2MA1217-0311                      & T7.5       & 14.19   (0.03)& 5  & 13.23   (0.02)& 5   & 13.34   (0.07)& 5   & 12.95   (0.18)& 5  &\phs0.96    &$-$0.12    & 0.39  & \nodata   \\
GJ570D                            & T7.5       & 13.80   (0.04)& 5  & 12.12   (0.02)& 5   & 12.77   (0.11)& 5   & 11.97   (0.07)& 5  &\phs1.68    &$-$0.64    & 0.80  & \nodata   \\
2MA0415-0935                      & T8.0       & 14.10   (0.03)& 5  & 12.29   (0.02)& 5   & 12.87   (0.07)& 5   & 12.11   (0.05)& 5  &\phs1.82    &$-$0.58    & 0.76  & \nodata\\
$\epsilon$ Ind BC\tablenotemark{a}& T1.0$+$T6.0& 9.97    (0.01)& 10 & 9.44    (0.02)& 10  & 9.39    (0.03)& 10  & 8.98    (0.04)& 10 &\phs0.53    &\phs0.04   & 0.41  & \nodata\\
\enddata
\tablecomments{(1) Source saturated in all four IRAC channels in 30-second FRAMETIME data, 2-second FRAMETIME data used instead,
(2) GTO program PID 33 target, 2-second FRAMETIME data,
(3) GTO program PID 33 target, 0.6-second FRAMETIME data,
(4) Channel 1 saturated for 30-second FRAMETIME data, used 2-second FRAMETIME data for this channel,
(5) Channel 2 saturated for 30-second FRAMETIME data, used 2-second FRAMETIME data for this channel,
(6) GTO program PID 33 target,
(7) Target source aperture possibly contaminated by flux from another nearby source,
(8) Target in crowded field, some contamination of source aperture by other nearby sources possible,
(9) {\it Spitzer} AOR target name incorrectly reads "2MA1410+1804",
(10) Wings of GJ1001 PSF may contaminate source aperture for target, however 2-second FRAMETIME data for channels 1 and 2 agree well with the 30-second FRAMETIME data,
(11) Strong nebulosity in background, especially in channel 4.}
\tablenotetext{a}{Known binary.}
\end{deluxetable}

\clearpage
\end{landscape}

\end{document}